\documentclass[aps, 12pt, preprint, preprintnumbers,  
aps, amsmath,amssymb,nofootinbib,superscriptaddress,hyperref
]{revtex4-1} 
\usepackage[utf8]{inputenc}
\usepackage{epsfig} 
\usepackage{wasysym} 
\usepackage{mathrsfs} 
\usepackage{graphicx} 
\usepackage{epstopdf}
\usepackage{amsfonts} 
\usepackage{amsbsy} 
\usepackage{amscd} 
\usepackage{pstricks} 
\usepackage{multirow}
\usepackage{slashed} 
\usepackage{tikz}
\usepackage{cancel}
\usetikzlibrary{decorations.pathmorphing,decorations.markings}

\newcommand{\beq}{\begin{equation}}
\newcommand{\eeq}{\end{equation}}
\newcommand{\bea}{\begin{eqnarray}}
\newcommand{\eea}{\end{eqnarray}}
\newcommand{\beas}{\begin{eqnarray*}}
\newcommand{\eeas}{\end{eqnarray*}}
\newcommand{\bi}{\begin{itemize}}
\newcommand{\ei}{\end{itemize}}

\def\tev{\,{\ifmmode\mathrm {TeV}\else TeV\fi}}
\def\gev{\,{\ifmmode\mathrm {GeV}\else GeV\fi}}
\def\to{\rightarrow}

\allowdisplaybreaks

\begin{document}
\begin{flushright}
\preprint{\textbf{OSU-HEP-18-06}}
\end{flushright}
\title{\Large Probing right handed neutrinos at the LHeC and lepton colliders using fat jet signatures}

\author{Arindam Das}
\email{\color{blue}{arindam@kias.re.kr}}
\affiliation{School of Physics, KIAS, Seoul 02455, Korea}

\author{Sudip Jana}
\email{\color{blue} {sudip.jana@okstate.edu}}
\affiliation{Department of Physics and Oklahoma Center for High Energy Physics, Oklahoma State University, Stillwater, OK 74078-3072, USA}

\author{Sanjoy Mandal}
\email{\color{blue}{smandal@imsc.res.in}}
\affiliation{The Institute of Mathematical Sciences,
C.I.T Campus, Taramani, Chennai 600 113, India}
\affiliation{Homi Bhabha National Institute, BARC Training School Complex,
Anushakti Nagar, Mumbai 400085, India}

\author{S. Nandi}
\email{\color{blue}{s.nandi@okstate.edu}}
\affiliation{Department of Physics and Oklahoma Center for High Energy Physics, Oklahoma State University, Stillwater, OK 74078-3072, USA}

\begin{abstract}

The inclusion of heavy neutral leptons (right-handed neutrinos) to the Standard Model (SM) particle content is one of the best motivated ways to account for the observed neutrino masses and flavor mixing. The modification of the charged and neutral currents from active-sterile mixing of the neutral leptons can provide novel signatures which can be tested at the future collider experiments. In this article, we explore the discovery prospect of a very heavy right handed neutrino to probe such extensions at the future collider experiments like Large Hadron electron Collider (LHeC) and linear collider. We consider the production of the heavy neutrino via the $t$ and $s$-channel processes and its subsequent decays into the semi-leptonic final states. We specifically focus on the scenario where the gauge boson produced from heavy neutrino decay is highly boosted, leading to a fat-jet. We study the bounds on the sterile neutrino properties from several past experiments and compare with our results.

\end{abstract}

\maketitle
\section{Introduction}
One of the most robust evidence that points out to an important inadequacy of the SM is the existence of the tiny but non-zero neutrino masses. It seems unlikely that the very small neutrino masses are generated by the same Higgs mechanism responsible for the masses of the other SM fermions due to the absence of right-handed neutrinos.  Even then,  extremely small Yukawa couplings, of the order of $\lesssim 10^{-12}$, must be invoked. There are various BSM extensions which have been proposed to explain small neutrino masses. Among those, one of the most appealing framework of light neutrino mass generation is the addition of new states that, once integrated out, generate the lepton  number violating dimension five Weinberg operator $\mathcal{O}_5=\frac{c}{\Lambda}LLHH$ \cite{Weinberg:1979sa} . This is embodied by the so-called seesaw mechanisms. There can be a few different variations of seesaw, Type-I \cite{typ1}, Type-II \cite{typ2}, Type-III \cite{typ3}, inverse \cite{inverse} and  radiative \cite{radiative} seesaw. 

Most of the UV completed seesaw models contain Standard Model (SM) gauge singlet heavy right handed neutrino $N$. Through the seesaw mechanism, the Majorana type Right Handed Neutrinos (RHNs) impart masses to the SM light neutrinos and hence establishes the fact that SM neutrinos have masses which have been experimentally observed in a several neutrino oscillation experiments \cite{numass}. These RHNs can have masses from eV scale to 10$^{14}$ GeV scale depending upon the models. For instance, the sterile neutrinos \cite{lightnu} with masses in the eV range could lead to effects in short distance neutrino oscillation experiments by introducing an additional mass squared difference, keV mass sterile neutrinos are potential candidates for “warm” dark matter, MeV scale sterile neutrinos can be possible explanation for MiniBoone \cite{mev} and there can be very heavy sterile neutrinos  with masses  $M_{GUT} \sim 10^{14}$ GeV, close to  $M_{GUT} \sim 10^{16}$ GeV in model of grand unified theories (GUTs). These RHNs, originally Standard Model (SM) gauge singlet, being mixed with the SM light neutrinos to interact with the SM gauge bosons. Depending on the mass of the gauge singlet RHNs and their mixings with the active neutrino states, seesaw mechanism can be tested at colliders \cite{delAguila:2008cj, Atre:2009rg,Datta:1993nm,Deppisch:2015qwa,Bhardwaj:2018lma,Das:2017gke,Abada:2018sfh,Jana:2018,Cottin:2018nms,Helo:2018qej,Accomando:2017qcs,Deppisch:2018eth,Das:2017pvt,Das:2017flq, Das:2017deo, Das:2016akd, Das:2017nvm,BhupalDev:2012zg, Das:2017zjc,Das:2017rsu, Chen:2013fna,Dev:2018upe,Dev:2017ftk,Cely:2012bz, Ibarra:2011xn, Ibarra:2010xw, Nemevsek:2018bbt,Nemevsek:2016enw,Maiezza:2015lza,Maiezza:2016bzp, Buchmuller:1990vh, Buchmuller:1991wn,Buchmuller:1992wm, Buchmuller:1991ce, Buchmuller:1991tu, Mondal:2016kof,Das:2014jxa, Chen:2011hc, Dev:2013wba, Dev:2015kca, Biswal:2017nfl, Das:2016hof, Das:2015toa},  as  well  as,  in other  non-collider  experiments,  such  as,  neutrinoless  double  beta  decay \cite{Dev:2013vxa, Mitra:2011qr,Dev:2014xea,Rodejohann:2012xd,Pas:2015eia,Gonzalez:2017mcg,Das:2017hmg,Das:2017nvm},  neutrino experiments \cite{lightnu,mev,nuex}, rare-meson decays \cite{Ali:2001gsa,Mandal:2016hpr,Mandal:2017tab}, muon $g-2$ \cite{g2},  lepton flavor violating processes $l_i \to l_j \gamma, \mu \to 3e , \mu \to e$ conversion in nuclei  \cite{Abada:2007ux,Abada:2008ea,Dinh:2012bp}, non-unitarity \cite{Antusch:2014woa,Antusch:2006vwa,Blennow:2016jkn,Fernandez-Martinez:2016lgt,Fernandez-Martinez:2015hxa}, etc.

We are specifically interested in the RHNs at the TeV scale so that they can be tested at the high energy colliders. At the LHC, the production cross section of the RHN decreases as the mass of RHN increases as a result of the properties of the constituent quarks of the proton beams. In the linear collider the electron and positron are collided to produce the RHN in association with a light neutrino through the dominant t-channel process.  A subdominant s-channel process also contributes \cite{ar1,ar2}. Otherwise a variety of RHN productions at the linear collider have been discussed in \cite{ar3} followed by the bounds on the light heavy mixing angles for the electron flavor at the linear collider with $500$ GeV and $1$ TeV collider energies. The low mass range of the RHN has been studied in \cite{ar4} which also predicts the limit on the light heavy mixing and the mass of the RHN up to a mass of $250$ GeV. The sterile neutrinos at the circular lepton colliders have been studied in \cite{ar5} which deals with a comprehensive discussion on the detectors from experimental point of view. Higgs searches from RHN has been studied in \cite{ar6} where the RHN has been produced from the $W$ and $Z$ mediated processes. Such a RHN decays into a Higgs and SM light neutrino and the Higgs can dominantly decay into a pair of $b$-quarks. Hence a $2b$ plus missing momentum will be a signal from this process. In this paper the RHN up to a $500$ GeV mass have been tested where the maximum center of mass energy is also taken up to $500$ GeV. The distinct and interesting signature of the RHN can be displaced vertex search if the mixing between the light and heavy neutrinos become extremely small. Such a scenario has been tested in \cite{ar7} for the colliders $240$ GeV, $350$ GeV and $500$ GeV. Another interesting work on the RHNs has been found in the form of \cite{ar8} where a variety of the colliders have been considered to test the observability of the RHN production. They have discussed several production modes of the RHNs at the LHC, lepton-Hadron collider (LHeC) \footnote{ In such a collider we can also nicely study the long lived particles in \cite{Curtin:2017bxr}, beyond the SM physics in \cite{Azuelos:2018syu}, leptoquarks \cite{AbelleiraFernandez:2012cc}, left-right model \cite{Lindner:2016lxq}, charged Higgs \cite{Azuelos:2017dqw} and heavy Majorana neutrinos \cite{Duarte:2018xst}. The LHeC design report can be found in \cite{AbelleiraFernandez:2012cc}.} \cite{Mandal:2018qpg} and linear collider. They have studied all possible modes of the RHN production in these colliders and compared the bounds on the light-heavy neutrino mixing angles. In the linear collider, the references \cite{ar4,ar5,ar6,ar7,ar8} did not go further than $500$ GeV as they constrained themselves within the center of mass energy of $500$ GeV. However, none of these papers studied the boosted object at the LHeC and linear collider respectively.

In our analysis we consider the following things:
\begin{enumerate}

\item{We study the prospect of discovery of RHNs at LHeC considering the boosted objects for the first time. In the LHeC we concentrate on the lepton number violating (LNV) and lepton number conserving (LNC) channels to produce the RHN in association with a jet $(j_1)$. Hence the RHN will decay into the dominant $\ell W$ and the $W$ will decay into a pair of jets. The daughter $W$ coming from the heavy RHN will be boosted and its hadronic decay products, jets, of the $W$ will be collimated such that they can form a fat jet $(J)$.Hence a signal sample of $\ell+j_1+J$ can be studied thoroughly at this collider. In this process people have mostly studied the lepton number conserving channel where as the lepton number violating will be potentially background free.  However, for clarity we study the combined channel and the corresponding SM backgrounds. We consider two scenarios at the LHeC where the electron and proton beams will have $60$ GeV and $7$ TeV energies where the center of mass energy becomes $\sqrt{s}=1.3$ TeV. We have also considered another center of mass energy at the $\sqrt{s}=1.8$ TeV where the proton beam energy is raised up to the $13.5$ TeV. For both of the colliders we consider the luminosity at $1$ ab$^{-1}$. Here the RHN is a first generation RHN $(N_1)$ and $\ell$ is electron $(e)$. Finally we study up to $3$ ab$^{-1}$ luminosity.}

\item{At the linear collider the production of the RHNs is occurring from the $s$- and $t$- channel processes in association with a SM light neutrino $(\nu)$. We consider the linear collider at two different center of mass energies, such as $\sqrt{s}=1$ TeV and $\sqrt{s}=3$ TeV which can probe up to a high mass of the RHNs such as $900$ GeV (at the $1$ TeV linear collider) and $2.9$ TeV (at the $3$ TeV linear collider) due to the almost constant cross section for the $N\nu$ production. For both of the center of mass energies we consider $1$ ab$^{-1}$ luminosity. Finally we study up to $3$ $(5)$ ab$^{-1}$ luminosity for the $1$ $(3)$ TeV linear collider.}

At this mass scale, the RHNs will be produced at rest, however, the daughter particles can be sufficiently boosted. We consider $N \to \ell W, W \to j j$ and $N\to h \nu, h \to b \overline{b}$ modes at the linear collider where $h$ is the SM Higgs boson. If the RHN is sufficiently heavy, such the, $M_N \geq 400$ GeV, the $W$ and $h$ can be boosted because $M_W$ and $M_h << \frac{M_N}{2}$. As a result $W$ and $h$ will produce a fat jet $(J)$ and a fat $b$ jet $(J_b)$ respectively. Therefore the signal will be $\ell+J$ plus missing momentum  and $J_b$ plus missing momentum in the $W$ and $h$ modes respectively at the linear collider. Therefore studying the signals and the backgrounds for each process we put the bounds in the mass- mixing plane of the RHNs.

\item{We want to comment that studying $e^- e^+ \to N_2 \nu_{\mu}/ N_3 \nu_{\tau}$ mode in the $Z$ mediated $s$-channel will be interesting where $N_2 (N_3)$ will be the second (third) generation RHN. Studying the signal events and the corresponding SM backgrounds one can also calculate the limits on the mixing angles involved in these processes. Such a process will be proportional to $\mid V_{\mu N} \mid^2 (\mid V_{\tau N} \mid^2)$. In these processes the signal will be $\mu (\tau)+j j$ plus missing momentum followed by the decay of $N_2 (N_3) \to \mu jj (\tau jj)$. One can also  calculate the bounds on the mass-mixing plane for different significances. A boosted analysis could be interesting, however, a non-boosted study might be more useful as the cross-section goes down with the rise in collider energy in these processes. Such signals can also be studied if the RHNs can decay through the LFV modes, such as $e^- e^+ \to N \nu_{e}, N \to \mu W, W\to j j$, however, $\mu \to e \gamma$ process will make this process highly constrained due to the strong limit $Br(\mu^+\to e^+\gamma) < 4.2\times 10^{-13}$ at the $90\%$ C. L. \cite{TheMEG:2016wtm}. The corresponding limits on $\tau$ are weaker \cite{Aubert:2009ag, OLeary:2010hau}. Such final states have been studied in \cite{ar1} for $M_N=150$ GeV, a high mass test with using boosted object will be interesting in future. A comprehensive LHC study has been performed in \cite{Antusch:2018bgr}.}

\item{The RHN produced at the linear collider may decay in to another interesting mode, namely, $N \to Z \nu, Z \to b \bar{b}$. Which can be another interesting channel where boosted objects can be stated. However, precision measurements at the $Z$-boson resonance using electron-positron colliding beams at LEP experiment strongly constrains Z boson current, and hence, $Zb\bar{b}$ coupling. This channel also suffers from larger QCD background compared to the leptonic decay of Z boson, and hence, leptonic decay of Z boson has better discovery prospect for this particular mode of RHN decay. On the other hand, SM Higgs , $h$, mostly decays ($\sim 60\%$) to $b\bar{b}$ due to large $hb\bar{b}$ coupling. Due to this, we focus on the Higgs decay mode of RHN, $N \to h \nu, h \to b \bar{b}$ to study the fat jet signature.
For the time being, we mainly focus on the first two items. The investigation of the mode, $N \to Z \nu, Z \to b \bar{b}$ is beyond the scope of this article and shall be presented in future work in detail.}

\end{enumerate}

The paper is organised as follows. in Sec. \ref{sec2}, we discuss the model and the interactions of the heavy neutrino with SM particles and also calculate the production cross sections at different colliders. In Sec.\ref{sec3} we discuss the complete collider study. 
In Sec.~\ref{sec4} we calculate the bounds on the mixing angles and compare them with the existing results. Finally, we conclude in Sec. \ref{sec5}.

\section{Model and the production mode}
\label{sec2}

In type-I seesaw \cite{typ1},  SM gauge-singlet right handed Majorana neutrinos $N_R^{\beta}$ are introduced,    
 where $\beta$ is the flavor index. $N_R^{\beta}$ have direct coupling with SM lepton doublets $\ell_{L}^{\alpha}$ and the SM Higgs doublet $H$.
The relevant part of the Lagrangian can be written as :
\bea
\mathcal{L} \supset -Y_D^{\alpha\beta} \overline{\ell_L^{\alpha}}H N_R^{\beta} 
                   -\frac{1}{2} M_N^{\alpha \beta} \overline{N_R^{\alpha C}} N_R^{\beta}  + H. c. .
\label{typeI}
\eea
After the spontaneous EW symmetry breaking
   by getting the vacuum expectation value (VEV) of the Higgs field, 
   $ H =\begin{pmatrix} \frac{v}{\sqrt{2}} \\  0 \end{pmatrix}$, 
    we obtain the Dirac mass matrix as $M_{D}= \frac{Y_D v}{\sqrt{2}}$.
Using the Dirac and Majorana mass matrices, the neutrino mass matrix can be written as 
\bea
M_{\nu}=\begin{pmatrix}
0&&M_{D}\\
M_{D}^{T}&&M_N
\end{pmatrix}.
\label{typeInu}
\eea
After diagonalizing this matrix, we obtain the seesaw formula for the light Majorana neutrinos as 
\bea
m_{\nu} \simeq - M_{D} M_N^{-1} M_{D}^{T}.
\label{seesawI}
\eea
For $M_N\sim 100$ GeV, we may find $Y_{D} \sim 10^{-6}$  with $m_{\nu}\sim 0.1$ eV.
However, in the general parameterization for the seesaw formula \cite{Casas:2001sr}, Dirac Yukawa term $Y_{D}$ can be as large as $1$, and this scenario is considered in this paper.
  
There is another seesaw mechanism, so-called inverse seesaw \cite{inverse}, where the light Majorana neutrino mass is generated through tiny lepton number violation.
The relevant part of the Lagrangian is given by
\bea
\mathcal{L} \supset - Y_D^{\alpha\beta} \overline{\ell_L^{\alpha}} H N_R^{\beta}- M_N^{\alpha \beta} \overline{S_L^{\alpha}} N_R^{\beta} -\frac{1}{2} \mu_{\alpha \beta} \overline{S_L^{\alpha}}S_L^{\beta^{C}} + H. c. ,
\label{InvYuk}
\eea 
where  $M_N$ is the Dirac mass matrix, $N_R^{\alpha}$ and $S_L^{\beta}$ are two SM-singlet heavy neutrinos with the same lepton numbers, and
   $\mu$ is a small lepton number violating Majorana mass matrix.
After the electroweak symmetry breaking the neutrino mass matrix is obtained as 
\bea
M_{\nu}=\begin{pmatrix}
0&&M_{D}&&0\\
M_{D}^{T}&&0&&M_N^{T}\\
0&&M_N&&\mu
\end{pmatrix}.
\label{InvMat}
\eea
After diagonalizing this mass matrix, we obtain the light neutrino mass matrix 
\bea
M_{\nu} \simeq M_{D} M_N^{-1}\mu M_N^{-1^{T}} M_{D}^{T}.
\label{numass}
\eea
Note that the small lepton number violating term $\mu$ is responsible for the tiny neutrino mass generation. The smallness of $\mu$ allows the $M_{D}M_N^{-1}$ parameter  to be order one even for an EW scale heavy neutrino. Since the scale of $\mu$ is much smaller than the scale of $M_N$, the heavy neutrinos become the pseudo-Dirac particles. This is the main difference between the type-I and the inverse seesaw.

Assuming $M_D M_N^{-1} \ll 1$, the flavor eigenstates ($\nu$) of  the light Majorana neutrinos can be expressed in terms of 
 the mass eigenstates of the light ($\nu_m$)  and heavy ($N_m$) Majorana neutrinos such as 
\bea 
  \nu \simeq {\cal N} \nu_m  + {\cal R} N_m,  
\eea 
where 
\bea
 {\cal R} = M_D M_N^{-1}, \; 
 {\cal N} =  \left(1 - \frac{1}{2} \epsilon \right) U_{\rm MNS} 
\eea
 with $\epsilon = {\cal R}^* {\cal R}^T$,  and $U_{MNS}$ is the usual neutrino mixing matrix 
 by which the mass matrix $m_\nu$ is diagonalized as  
\bea
   U_{MNS}^T m_\nu U_{MNS} = {\rm diag}(m_1, m_2, m_3). 
\eea
In the presence of $\epsilon$, the mixing matrix ${\cal N}$ is not unitary \cite{Antusch:2006vwa,Abada:2007ux,Antusch:2014woa, Antusch:2016brq}. 
Considering the mass eigenstates, the charged current interaction in the Standard Model is given by 
\bea 
\mathcal{L}_{CC}= 
 -\frac{g}{\sqrt{2}} W_{\mu}
  \bar{e} \gamma^{\mu} P_L 
   \left( {\cal N} \nu_m+ {\cal R} N_m \right) + h.c., 
\label{CC}
\eea
where $e$ denotes the three generations of the charged 
 leptons in the vector form, and 
$P_L =\frac{1}{2} (1- \gamma_5)$ is the projection operator. 
Similarly, the neutral current interaction is given by 
\bea 
\mathcal{L}_{NC}= 
 -\frac{g}{2 c_w}  Z_{\mu} 
\left[ 
  \overline{\nu_m} \gamma^{\mu} P_L ({\cal N}^\dagger {\cal N}) \nu_m 
 +  \overline{N_m} \gamma^{\mu} P_L ({\cal R}^\dagger {\cal R}) N_m 
+ \left\{ 
  \overline{\nu_m} \gamma^{\mu} P_L ({\cal N}^\dagger  {\cal R}) N_m 
  + h.c. \right\} 
\right] , 
\label{NC}
\eea
 where $c_w=\cos \theta_w$ is the weak mixing angle. 
Because of non-unitarity of the matrix ${\cal N}$, 
 ${\cal N}^\dagger {\cal N} \neq 1$ and  
 the flavor-changing neutral current occurs. 

\begin{figure*}[h]
\begin{center}
\includegraphics[scale=0.4]{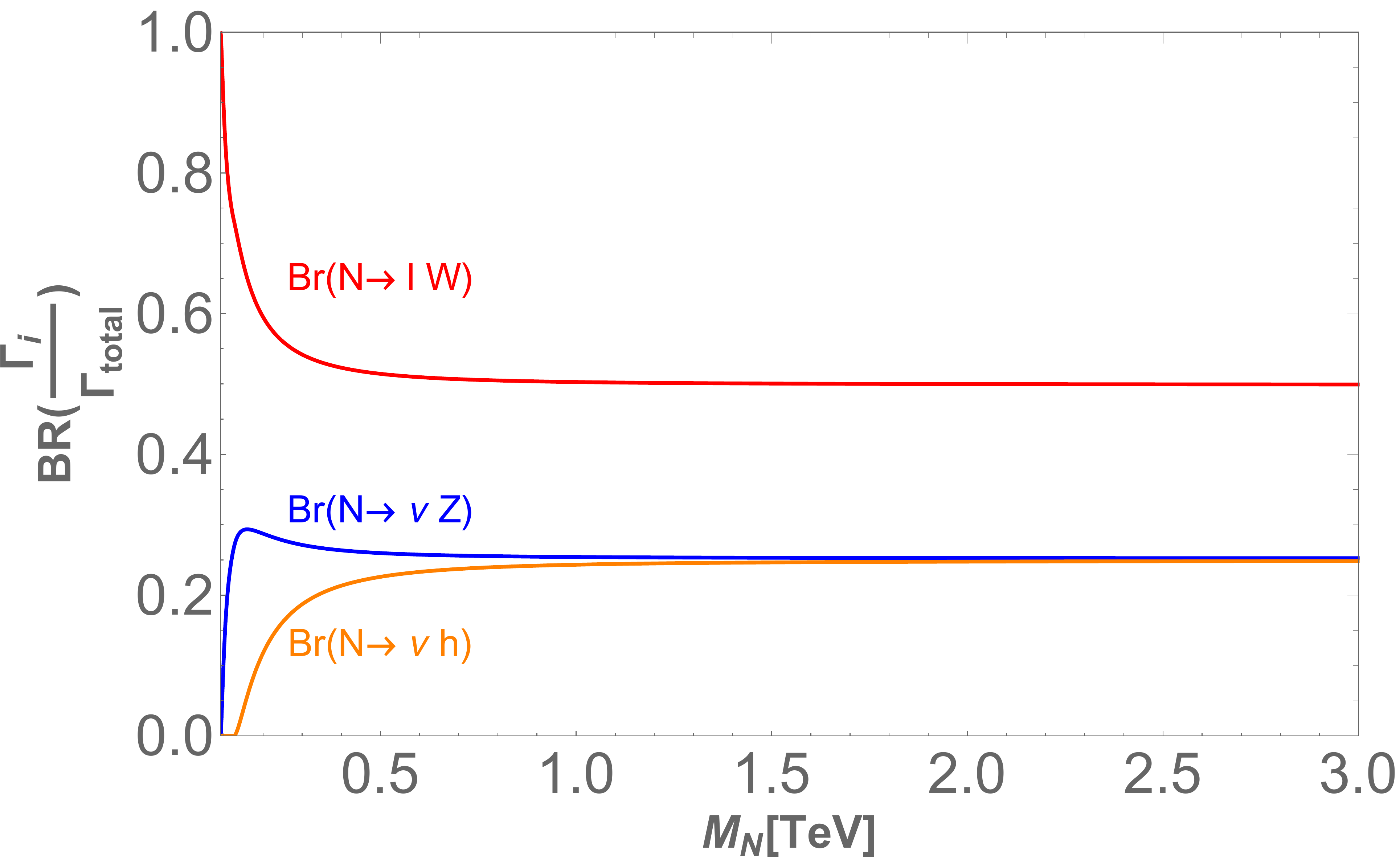}
\caption{Heavy neutrino branching ratios ($BR_i$) for different decay modes are shown with respect to the heavy neutrino mass $\left(M_N\right)$.}
\label{fig:BR}
\end{center}
\end{figure*}

  The dominant decay modes of the heavy neutrino are 
 $N \to \ell W$, $\nu_{\ell} Z$, $\nu_{\ell} h$ and the corresponding partial decay widths are respectively given by
\bea
\Gamma(N \rightarrow \ell W) 
 &=& \frac{g^2 |V_{\ell N}|^{2}}{64 \pi} 
 \frac{ (M_N^2 - M_W^2)^2 (M_N^2+2 M_W^2)}{M_N^3 M_W^2} ,
\nonumber \\
\Gamma(N \rightarrow \nu_\ell Z) 
 &=& \frac{g^2 |V_{\ell N}|^{2}}{128 \pi c_w^2} 
 \frac{ (M_N^2 - M_Z^2)^2 (M_N^2+2 M_Z^2)}{M_N^3 M_Z^2} ,
\nonumber \\
\Gamma(N \rightarrow \nu_\ell h) 
 &=& \frac{ |V_{\ell N}|^2 (M_N^2-M_h^2)^2}{32 \pi M_N} 
 \left( \frac{1}{v }\right)^2.
\label{widths}
\eea 
The decay width of heavy neutrino into charged gauge bosons being twice as large as neutral one owing to the two degrees of freedom $(W^{\pm})$.
We plot the branching ratios $BR_i \left(= {\Gamma_{i}}/{\Gamma_{\rm total}}\right)$ of the respective decay modes $\left(\Gamma_{i}\right)$ with respect to the total decay width $\left(\Gamma_{\rm total}\right)$ of the heavy neutrino into $W$, $Z$ and Higgs bosons in Fig.~\ref{fig:BR}
as a function of the heavy neutrino mass $\left(M_N\right)$. Note that for larger values of $M_N$, the branching ratios can be obtained as 
\bea
BR\left(N\rightarrow \ell W\right) : BR\left(N\rightarrow \nu Z\right) : BR\left(N\rightarrow \nu H\right) \simeq 2: 1: 1.
\eea 
\subsection{Production cross section at LHeC}
\begin{figure}[]
\centering
\includegraphics[width=0.5\textwidth]{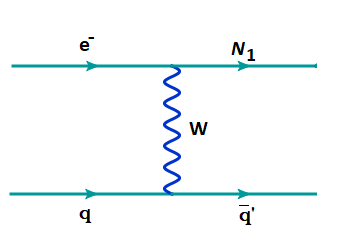}
\caption{Production process, $e p \to N_1 j_1$ ,of the RHN at the LHeC through a $t$ channel $W$ boson exchange}
\label{LHeC-F}
\end{figure}

The LHeC can produce the RHN in the process $e~p \to N_1 j_1$ through the $t$- channel exchanging the $W$ boson.
In this case the first generation RHN ($N_1$) will be produced. The corresponding Feynman diagram is given in Fig.~\ref{LHeC-F}. 
The total differential production cross section for this process is calculated as 
\bea
\frac{d \hat{\sigma}_{LHeC}}{d \cos\theta} &=&\frac{3.89\times 10^8}{32 \pi} 3 \times \frac{1}{3} \Big(\frac{1}{2}\Big)^2 \Big(\frac{M_{inv}^2 -M_N^2}{M_N^2}\Big) \times \nonumber \\
&& \frac{256 C_{\ell}^2 C_{q}^2  \Big(\frac{M_{inv}^2 -M_N^2}{4}\Big)} {\Big[M_N^2 -2\Big\{ \frac{M_{inv}^2}{4} \Big(1-\cos\theta\Big) \Big\}+ \frac{M_{inv}^2}{4} \Big(1+\cos\theta\Big)  \Big]^2+\Gamma_W^2 M_W^2 } 
\eea
where $C_\ell=C_q=\frac{g}{2^{\frac{3}{2}}}$.
\begin{figure}[]
\centering
\includegraphics[width=0.475\textwidth]{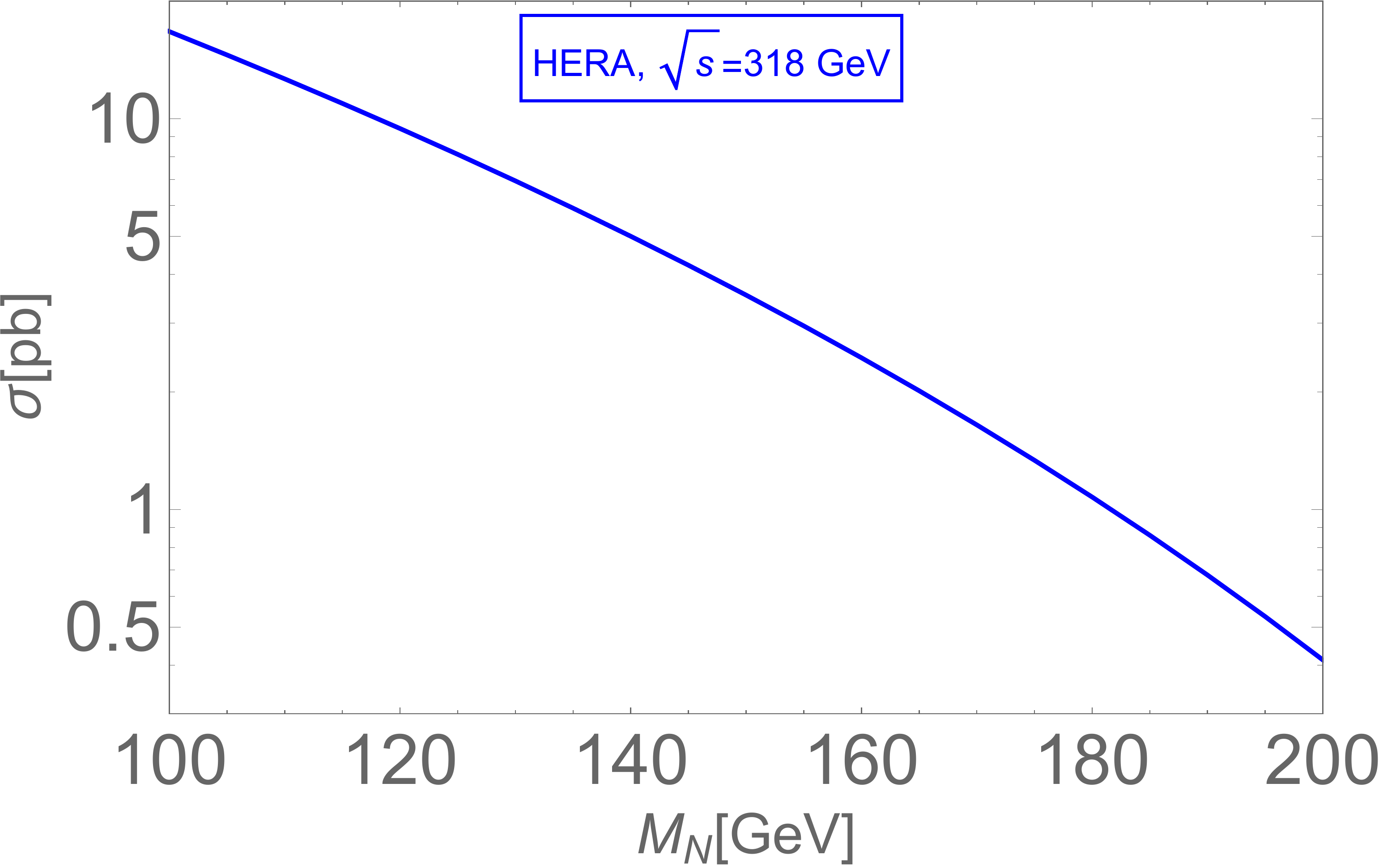}
\includegraphics[width=0.475\textwidth]{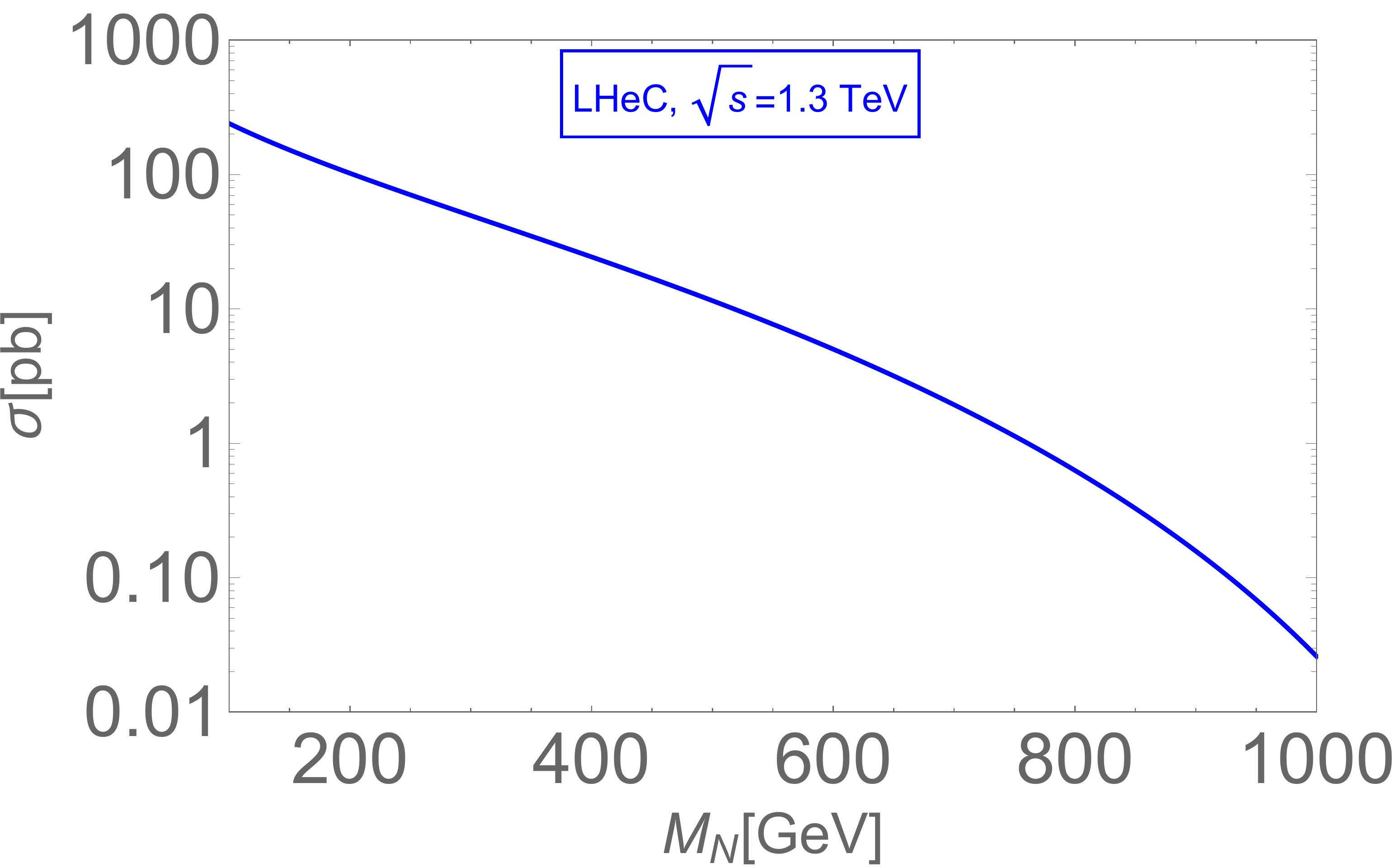}\\
\includegraphics[width=0.475\textwidth]{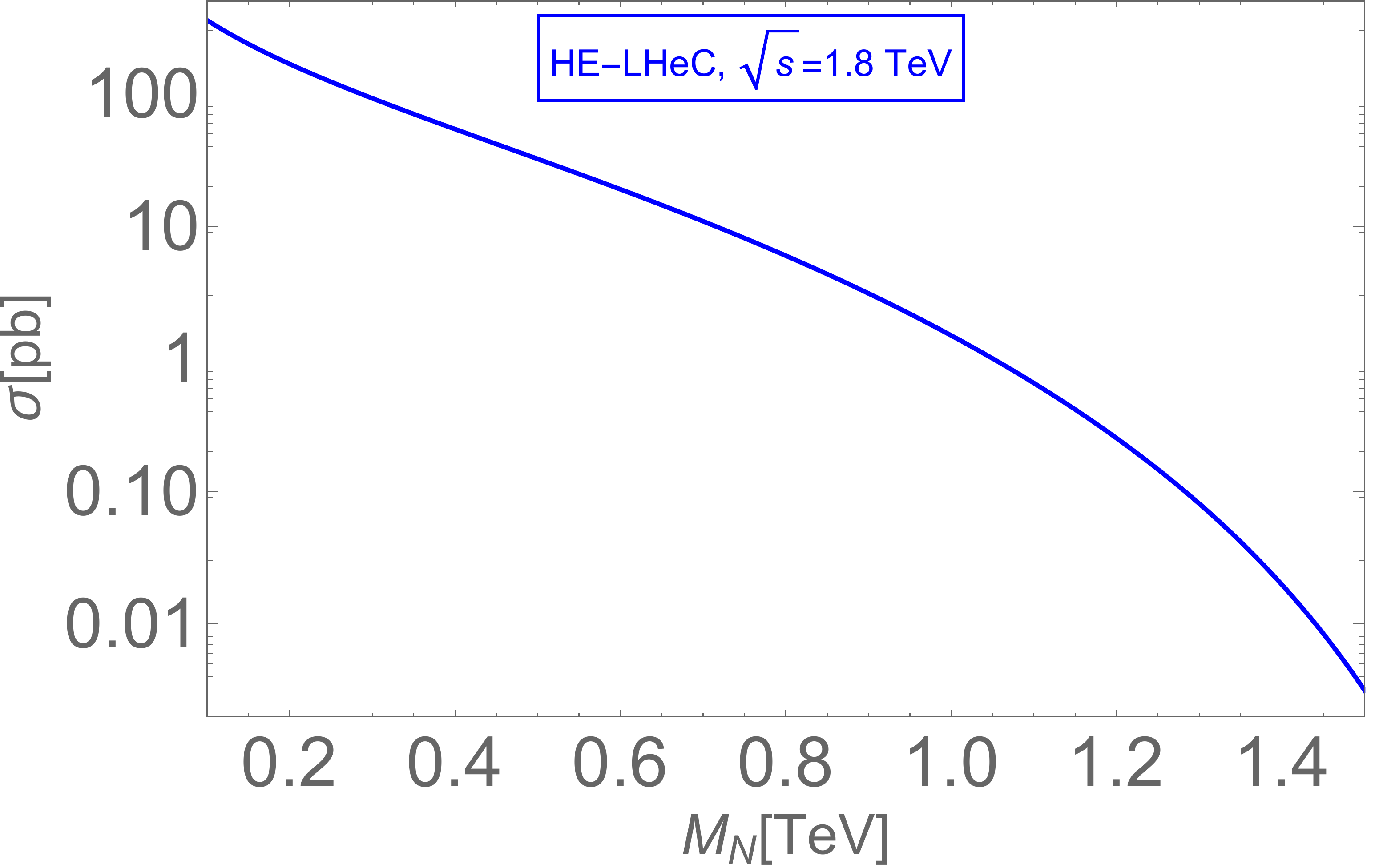}
\caption{RHN production cross section at the LHeC considering $e~p \to N_1 j$ process for the $e~p$ collider at $\sqrt{s}=318$ GeV (HERA, top left panel), $\sqrt{s}=1.3$ TeV (LHeC, top right panel) and $\sqrt{s}=1.8$ TeV (HE-LHeC, bottom panel).}
\label{LHeC00}
\end{figure}
Performing the integration over $\cos\theta$ between $[-1, 1]$ we find the cross section as $\hat{\sigma}_{LHeC}$ and finally convoluting the PDF (CTEQ5M) \cite{Pumplin:2002vw} we get the total cross section as 
\bea
\sigma= \sum_{i} \int^{1}_{\frac{M^2}{E_{CM}^2}} dx~q_i (x, \sqrt{x} E_{CM})~\hat{\sigma}_{LHeC}(\sqrt{x} E_{CM})
\eea
where $E_{CM}$ is the center of mass energy of the LHeC and $i$ runs over the quark flavors. For different center of mass energies $E$ will be different. In Fig.~\ref{LHeC00} we plot the total production cross sections of $N_1$ at the three different collider energies such as $\sqrt{s}=318$ GeV (HERA), $\sqrt{s}=1.3$ TeV (LHeC) and $\sqrt{s}=1.8$ TeV (High Energy LHeC (HE-LHeC)) respectively. The cross section in Fig.~\ref{LHeC00} is normalized by the square of the mixing to correspond the maximum value for a fixed $M_N$ according to the relevant part of the charged current interaction in Eq.~\ref{CC}.

\subsection{Production cross section at linear collider}
\begin{figure}[]
\centering
\includegraphics[width=1.00\textwidth]{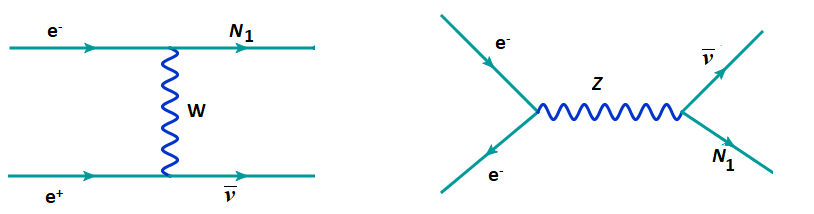}
\caption{RHN production processes at the linear collider. The left panel is the dominant $t$ channel process and the right panel is $s$ channel process to produce the $e^+ e^- \to N_1\nu_1$. To produce $N_2 \nu_2$ and $N_3 \nu_3$, the $Z$ mediated $s$ channel process will act.}
\label{ILC11}
\end{figure}
\begin{figure}[]
\centering
\includegraphics[width=0.475\textwidth]{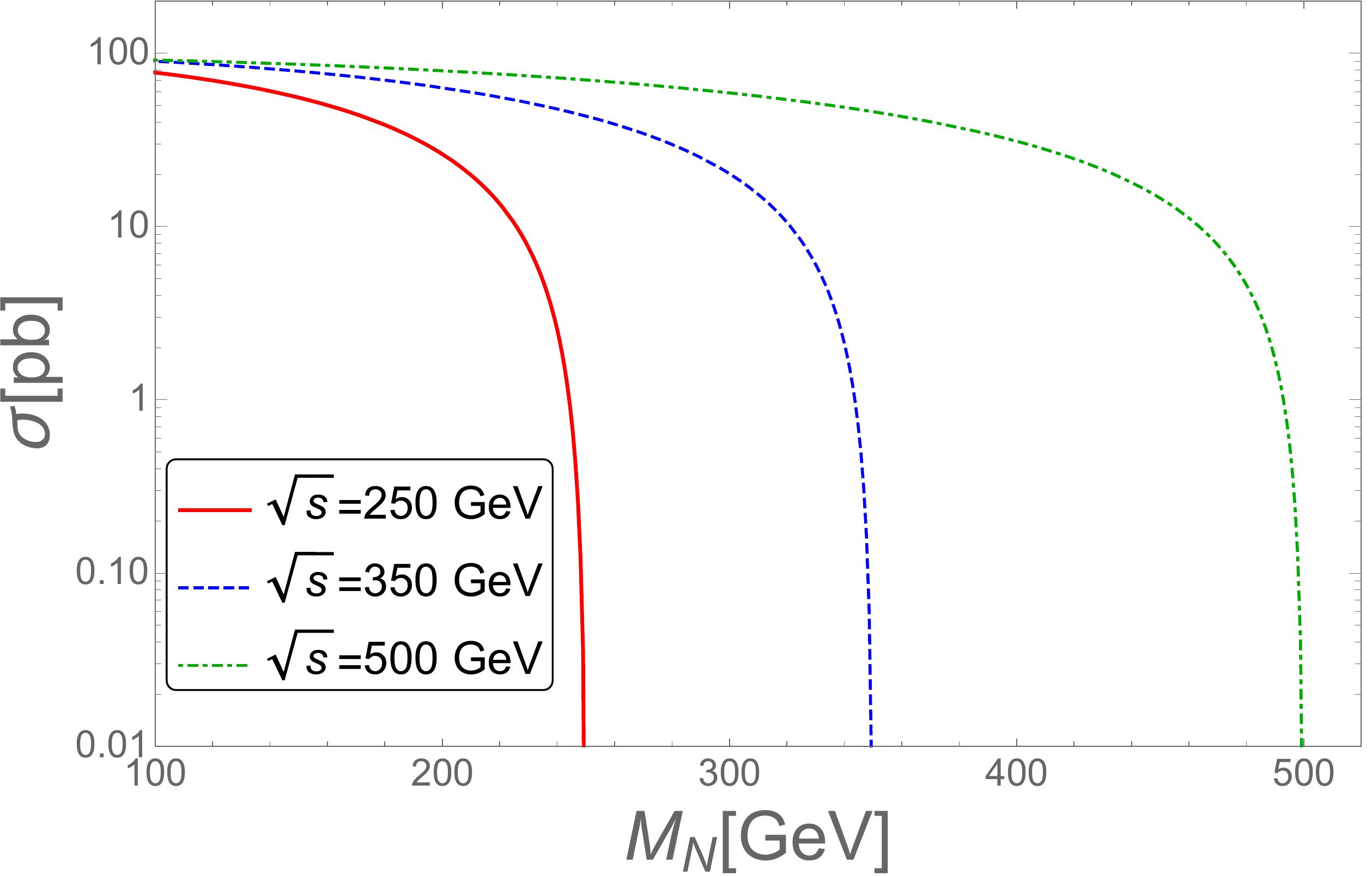}
\includegraphics[width=0.475\textwidth]{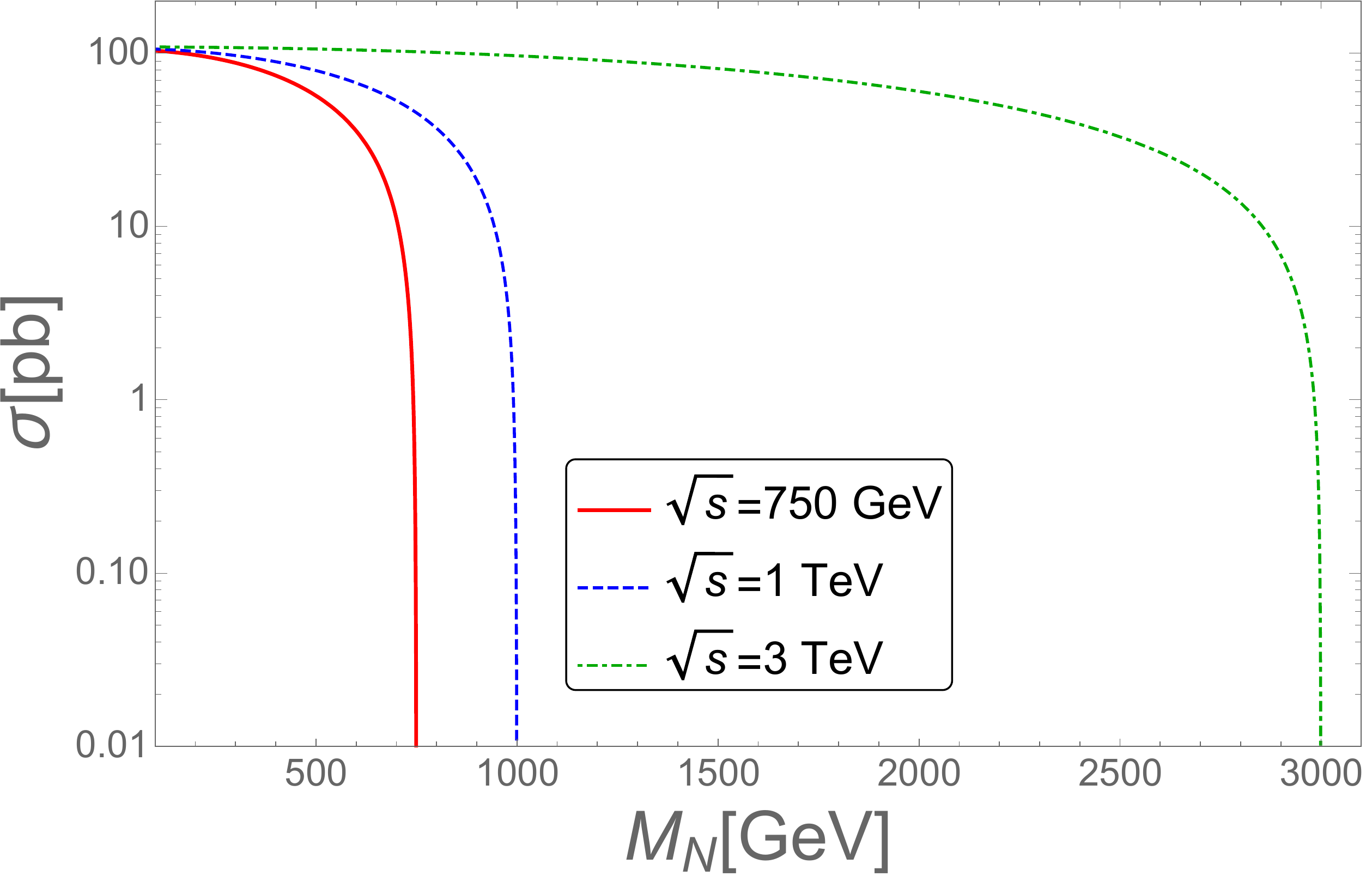}
\caption{RHN production cross section at the linear collider considering $e^{+} e^{-} \to N_1 \nu_1$ process at the different center of mass energies.}
\label{LC0}
\end{figure}

The linear collider can produce the heavy neutrino 
 in the process $e^+ e^- \to \overline{\nu_1} N_1$ 
 through $t$ and $s$-channels exchanging the $W$ and $Z$ bosons, respectively. 
 The corresponding Feynman diagrams are given in Fig.~\ref{ILC11}.
The total differential production cross section for this process is calculated as 
\bea
\frac{d \sigma_{ILC}}{d \cos\theta}  
&=& (3.89\times 10^{8} \; {\rm pb}) \times 
\frac{\beta}{32 \pi s} 
\frac{s + M_N^2}{s} 
\left( \frac{1}{2} \right)^2 
\nonumber \\
&\times& 
\left[   
\frac{16  C_1^2 C_2^2 \left( s^2 - M_N^4 \right) 
(1+\cos\theta) (1+ \beta \cos\theta)}
{(M_N^2 -\frac{s -M_N^2}{2} (1-\beta \cos\theta)- M_W^2)^{2}+ M_W^2 \Gamma_W^2} 
\right.
\nonumber \\
&+&
\frac{ \left( 
4 (C^{2}_{A_{e}}+C^{2}_{V_{e}})
(C^{2}_{A_{\nu}}+C^{2}_{V_{\nu}}) (1+\beta \cos^2 \theta )
+ 16 C_{A_{e}}C_{V_{e}}C_{A_{\nu}}C_{V_{\nu}} (1+\beta) \cos\theta 
\right)(s^2 - M_N^4)}{(s -M_Z^2)^2 + M_Z^2 \Gamma_Z^2} 
\nonumber \\
&-&
32 C^{2}_{1} C^{2}_{A_{e}} (s^2 - M_N^4 ) 
(1+ \cos\theta)(1+\beta \cos\theta) 
\nonumber \\
&&
\left.
 \times 
\frac{
\left(M_N^2 - \frac{s-M_N^2}{2}(1-\beta \cos\theta) - M_W^2 \right)(s - M_Z^2) 
+ M_W M_Z \Gamma_W \Gamma_Z  
}
{
((M_N^2 - \frac{s-M_N^2}{2} (1-\beta \cos\theta) -M_W^2)^{2} + M_W^2 \Gamma_W^2 )
 ( (s - M_Z^2)^2 + M_Z^2 \Gamma_Z^2 ) 
}
\right], 
\label{XILC}
\eea
where $\beta=(s-M_N^2)/(s+M_N^2)$,  
\bea
&&
C_{1}= -C_2 = \frac{g}{2 \sqrt{2}}, \; 
C_{A_{\nu}}= C_{V_{\nu}}= \frac{g}{4 \cos \theta_W}, 
\nonumber \\ 
&&
C_{A_{e}}= 
\frac{g}{2 \cos \theta_w} 
\left( 
-\frac{1}{2} + 2 \sin^2 \theta_w 
\right), \; 
C_{V_{e}}= - \frac{g}{4 \cos\theta_w}.
\eea
The total production cross section for the process 
 $e^+ e^- \to \overline{\nu_1} N_1$ 
 from the $t$ and $s$ channel processes at the linear collider at different center of mass energies are shown in Fig.~\ref{LC0}. 

\begin{figure}[]
\centering
\includegraphics[width=0.475\textwidth]{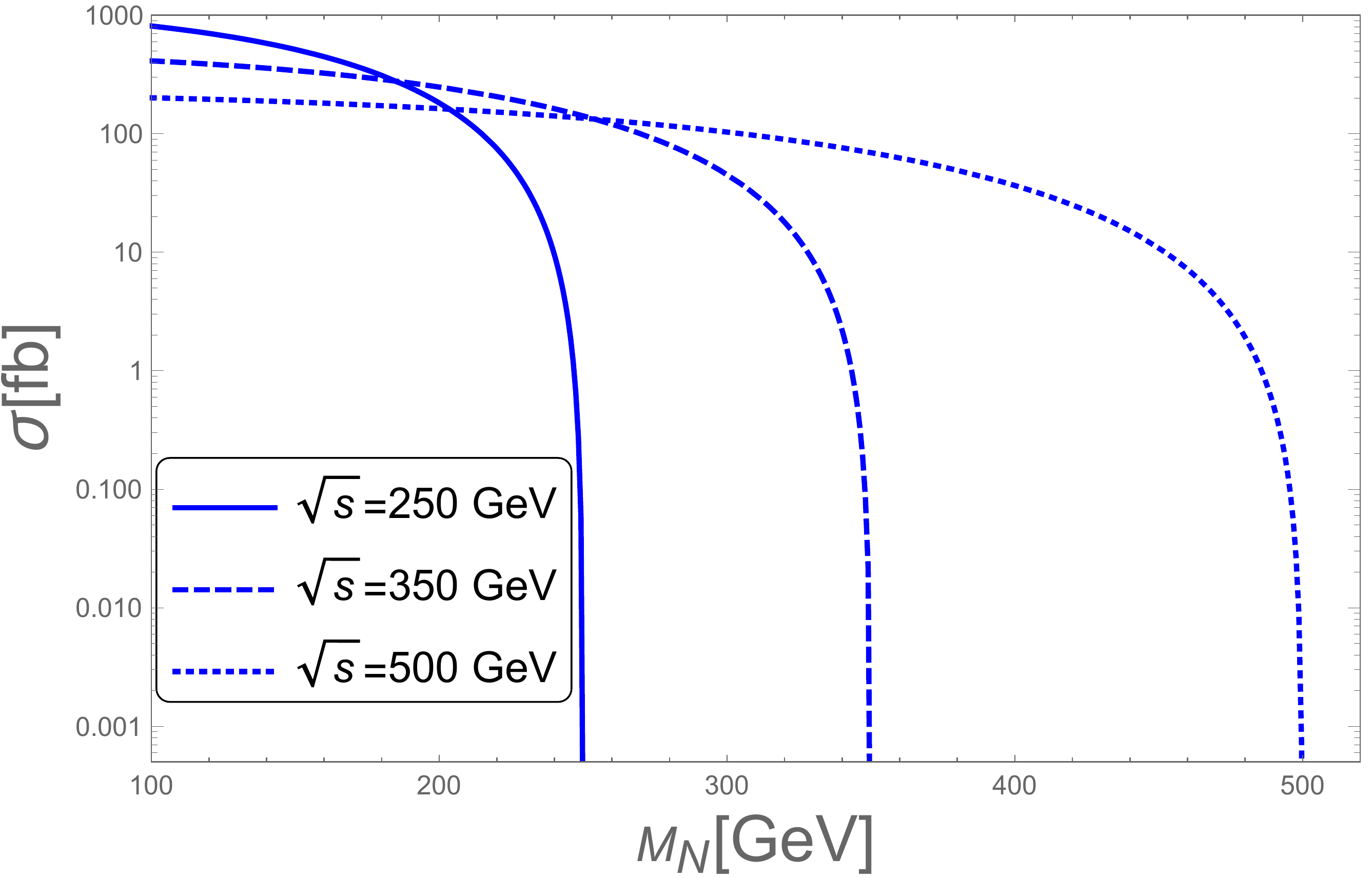}
\includegraphics[width=0.475\textwidth]{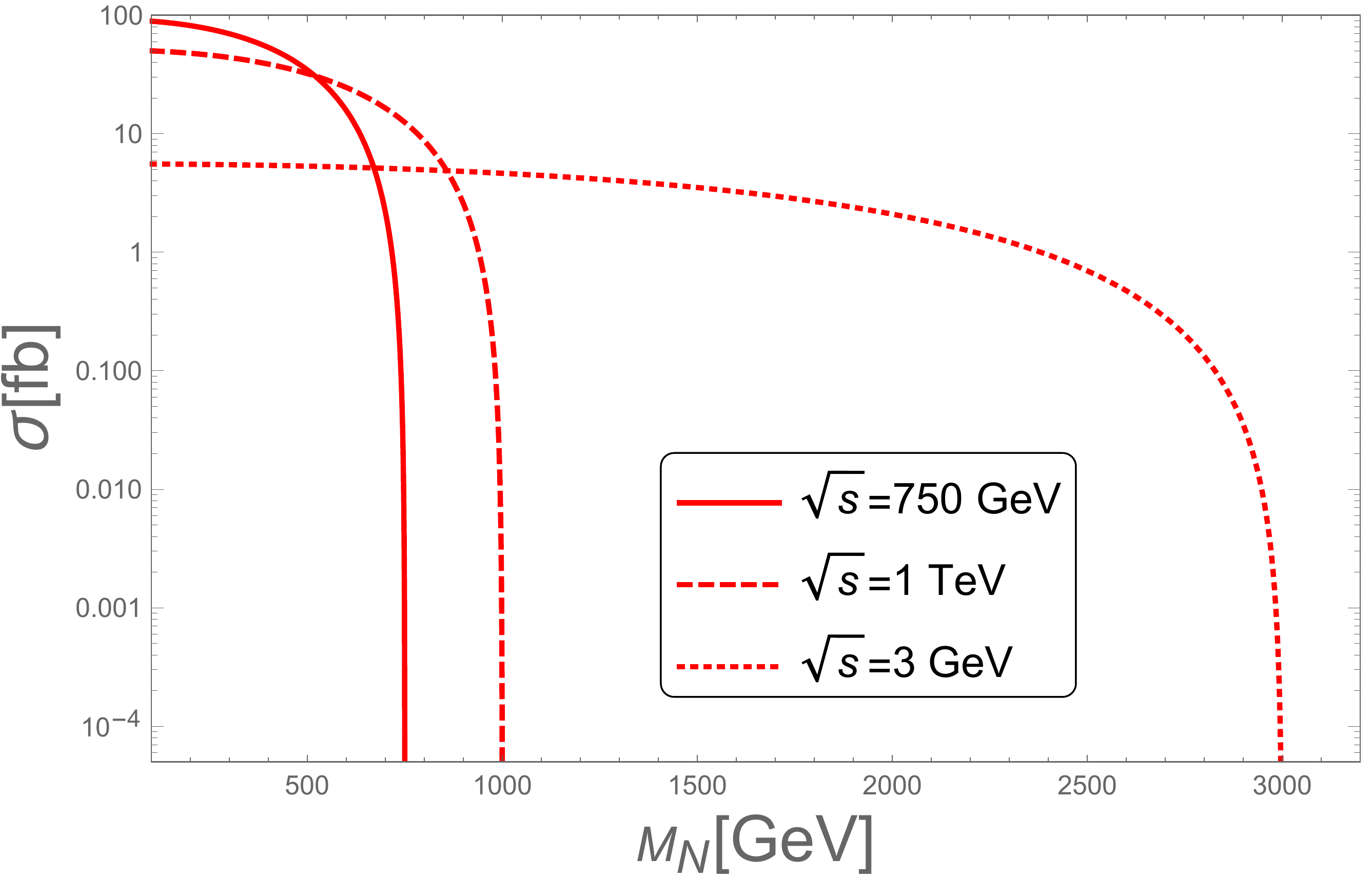}
\caption{RHN production cross section at the linear collider considering $e^+ e^- \to N_2 \nu_2~(N_3 \nu_3) $ process at the different center of mass energies from the $s$ channel $Z$ boson exchange.}
\label{LC00}
\end{figure}

The $s$ channel $Z$ mediated process can produce the second (third) generation of RHNs, $N_2 (N_3)$ in association with $\nu_2 (\nu_3)$. The cross sections for different center of mass energies have been given in Fig.~\ref{LC00}. The cross section in this mode decreases with the increase in the center of mass energy. Such modes can reach up to a cross section of $1$ pb for $M_N=100$ GeV at $\sqrt{s} =250$ GeV. Consider the leading decay mode of the RHN into $W$ and $\ell~(\mu, \tau)$ followed by the hadronic decay of the $W$ could be interesting to probe the corresponding mixing angles. 
The cross sections in Figs.~\ref{LC0} and \ref{LC00} are normalized by the square of the mixing to correspond the maximum value for a fixed $M_N$ according to the relevant part of the charged current and neutral current interactions in Eqs.~\ref{CC} and \ref{NC} respectively.
\section{Collider Analysis}
\label{sec3}
We implement our model in FeynRules \cite{Alloul:2013bka}, generate the UFO file of the model 
for MadGraph5-aMC@NLO \cite{Alwall:2014hca} to calculate the signals and the backgrounds.
Further we use PYTHIA6 \cite{Sjostrand:2006za} for LHeC as used in \cite{Mandal:2018qpg} and PYTHIA8 \cite{Sjostrand:2014zea} for the linear colliders, where subsequent decay, initial 
state radiation, final state radiation and hadronisation have been carried out. 
We have indicated in \cite{Das:2017gke, Bhardwaj:2018lma} that if the RHNs are sufficiently heavy, 
the daughter particles can be boosted. We prefer the hadronic decay mode of the $W$ where the jets 
can be collimated so that we can call it a fat-jet $(J)$. Such a topology is very powerful to discriminate the  signal from the SM backgrounds.
We perform the detector simulation using DELPHES version 3.4.1 \cite{deFavereau:2013fsa}. The detector card for the LHeC has been used from \cite{LHeC-card}. We use the ILD card for the linear collider. 
In our analysis the jets are reconstructed by Cambridge-Achen algorithm \cite{Dokshitzer:1997in, Wobisch:1998wt} implemented in Fastjet package \cite{Cacciari:2011ma,Cacciari:2005hq}  
with the radius parameter as $R=0.8$. 

\begin{figure}[]
\centering
\includegraphics[width=0.65\textwidth]{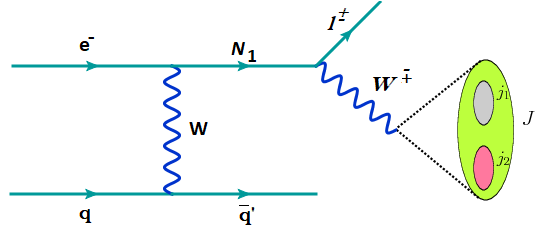}
\caption{$e+J+j_1$ final state at the LHeC and HE-LHeC.}
\label{LHeC010}
\end{figure}

\begin{figure}[]
\centering
\includegraphics[width=1.00\textwidth]{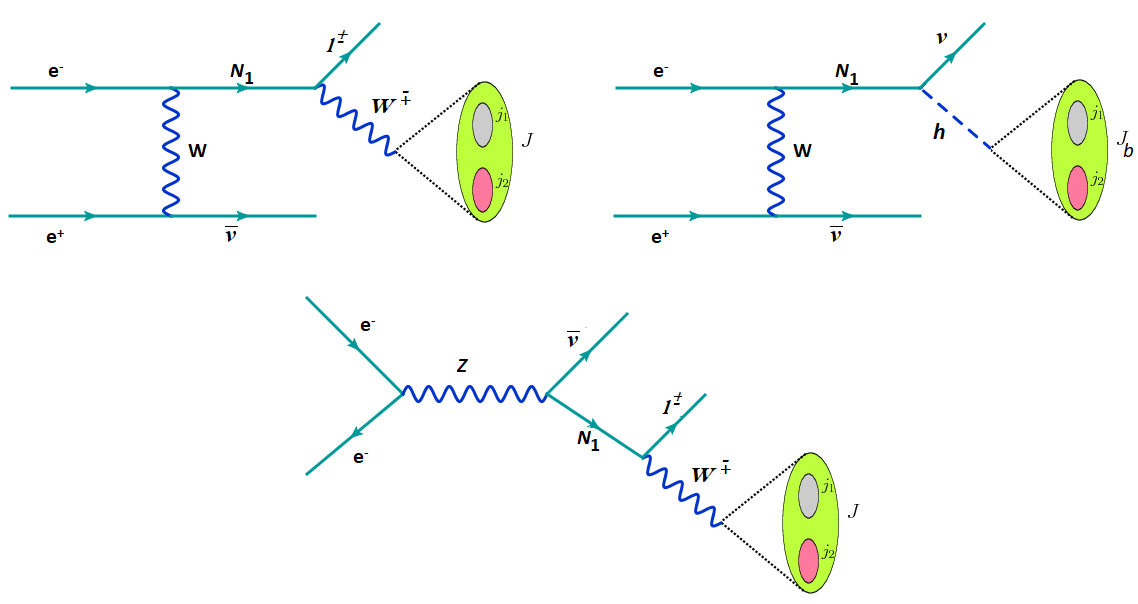}
\caption{$e+J+p_T^{miss}$ and $J_b+p_T^{miss}$ final states at the linear colliders.}
\label{ILC001}
\end{figure}

We study the production of the first generation RHN $(N_1)$ and its subsequent leading decay mode $(e~p \to N_1~j_1, N_1 \to W e, W \to J)$ at the LHeC with $\sqrt{s}=1.3$ TeV and $1.8$ TeV 
center of mass energies. The corresponding Feynman diagram is given in Fig.~\ref{LHeC010}. 
We also study the RHN production at the linear collider (International Linear Collider, ILC) at $\sqrt{s}=1$ TeV and CLIC at $\sqrt{s}=3$ TeV collider energies. However, for simplicity we will
use the term linear collider unanimously. At the linear collider we consider two sets of signals after the production of the RHN, such that, $ e^+ ~e^- \to N_1 ~\nu, N_1 \to W e, W \to J$ and $e^+ ~e^- \to N_1 ~\nu, N_1 \to h \nu, h \to J_b$ where $J_b$ is a fat $b$-jet coming from the boosted SM Higgs decay in the dominant mode. For the two types of colliders we consider $1000$ fb$^{-1}$ luminosity. The corresponding Feynman diagrams are given in Fig.~\ref{ILC001}.
For the analysis of signal and background events we use the following set of basic cuts,\\
\begin{enumerate}
\item
Electrons in the final state should have the following transverse momentum $(p_T^e)$ and pseudo-rapidity $(|\eta^e|)$ as $p_{T}^{e}>10$ GeV, $|\eta^{e}|<2.5.$
\item
Jets are ordered in $p_{T}$, jets should have $p_{T}^{j}>10$ GeV and $|\eta^{j}|<2.5$.
\item
Photons are counted if $p_{T}^{\gamma}>10$ GeV and $|\eta^{\gamma}|<2.5$.
\item
 Leptons should be separated by $\Delta R_{\ell\ell}>0.2$.
 \item
 The leptons and photons are separated by $\Delta R_{\ell\gamma}>0.3$.
\item
 The jets and leptons should be separated by $\Delta R_{\ell j}>0.3$.
\item
Fat Jet is constructed with radius parameter $R=0.8$.
\end{enumerate}
\begin{figure}[]
\centering
\includegraphics[width=0.475\textwidth]{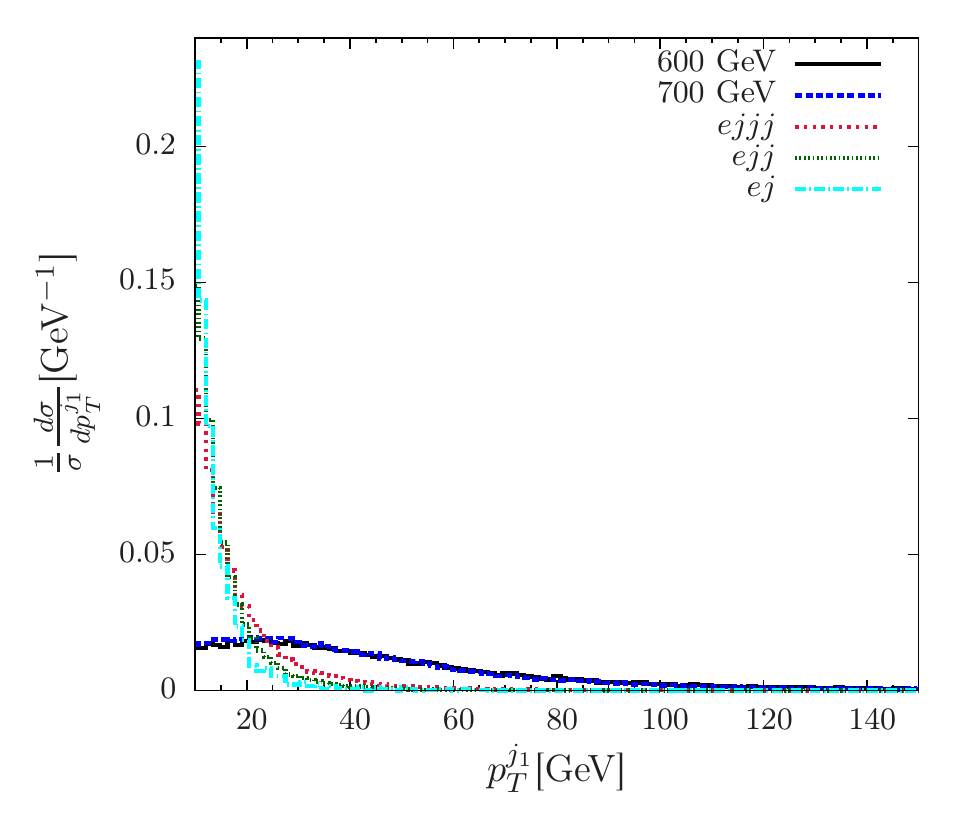}
\includegraphics[width=0.475\textwidth]{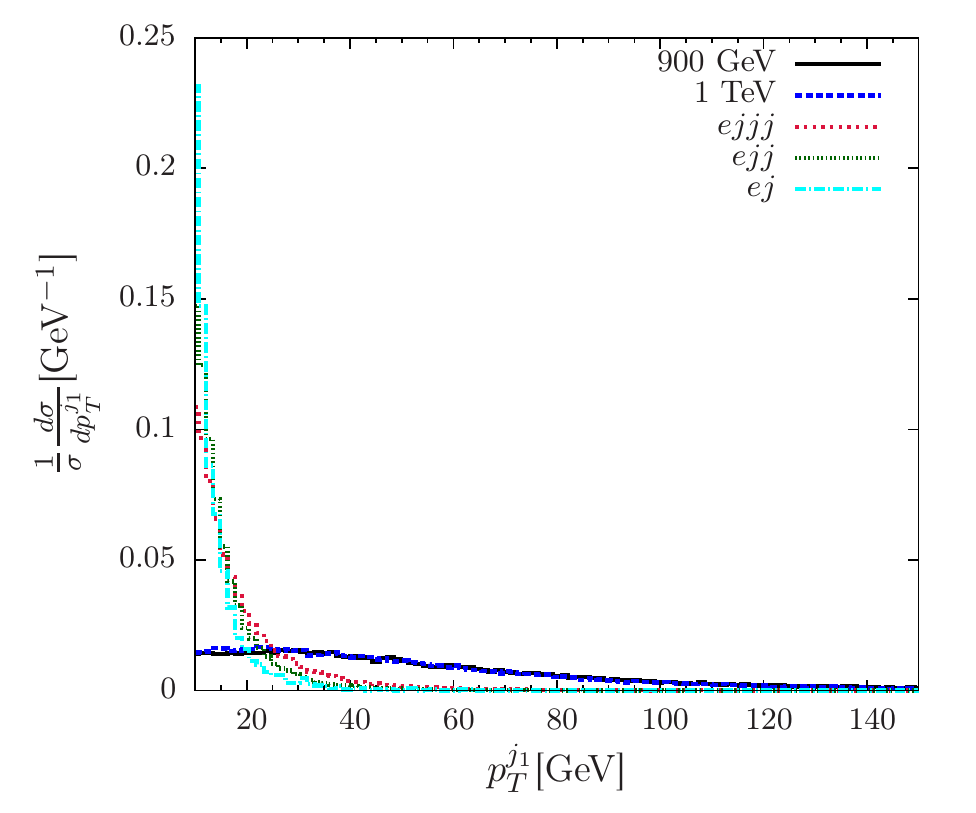}
\caption{Transverse momentum distribution of the associated jet $(p_T^{j_1})$ from the signal and background events for $M_N=600$ GeV and $700$ GeV at the $\sqrt{s}=1.3$ TeV LHeC (left panel) and $M_N=900$ GeV and $1$ TeV at the $\sqrt{s}=1.8$ TeV LHeC (right panel)}
\label{LHeCfig0}
\end{figure}

\subsection{LHeC analysis for the signal $e^{-}p\to j N_{1}\to e^{\pm}+J+j_1$}
Producing $N_1$ at the LHeC and followed by its decay into leading mode to study the boosted objects, we consider the final state $ e^{\pm}+J+j_1$.
In this case we have two different processes, one is them is the $e^++J+j_1$ and the other one is $e^-+J+j_1$. The first one is the Lepton Number Violating (LNV) channel 
and the second one is the Lepton Number Conserving (LNC). At the time of showing the results we combine LNV and LNC channels to obtain the final state as $e^\pm+J+j_1$.

\begin{figure}[]
\centering
\includegraphics[width=0.475\textwidth]{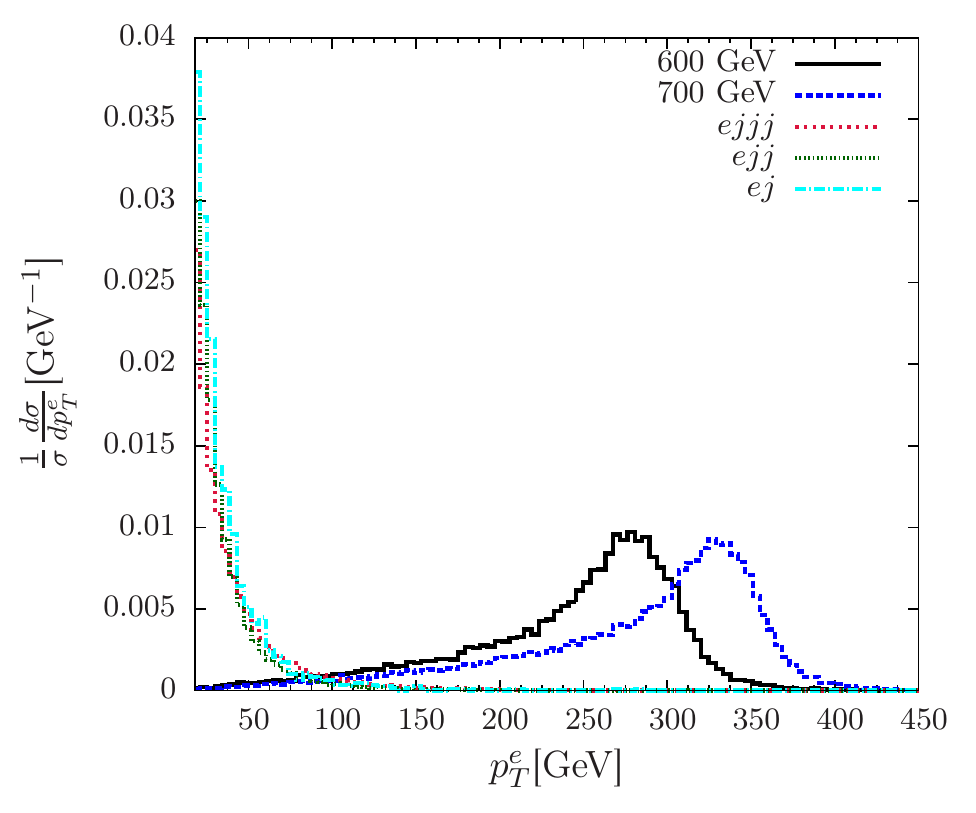}
\includegraphics[width=0.475\textwidth]{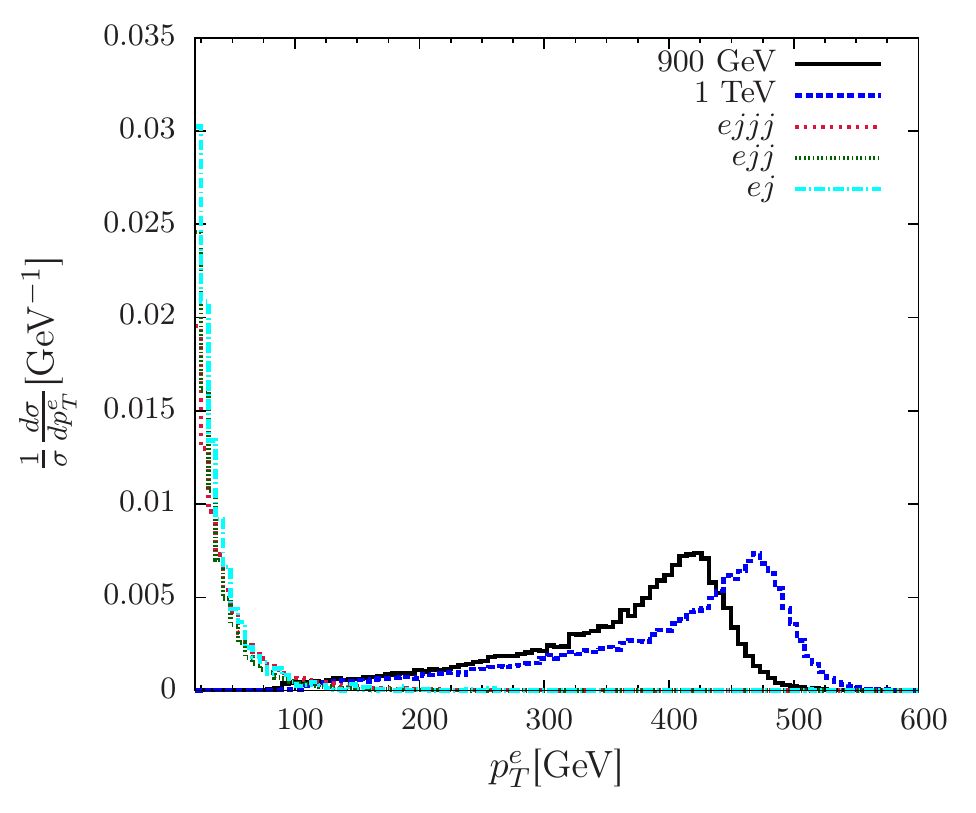}
\caption{Transverse momentum distribution of the electron $(p_T^e)$ from the signal and background events for $M_N=600$ GeV and $700$ GeV at the $\sqrt{s}=1.3$ TeV LHeC (left panel) and $M_N=900$ GeV and $1$ TeV at the $\sqrt{s}=1.8$ TeV HE-LHeC (right panel)}
\label{LHeCfig1}
\end{figure}

\begin{figure}[]
\centering
\includegraphics[width=0.475\textwidth]{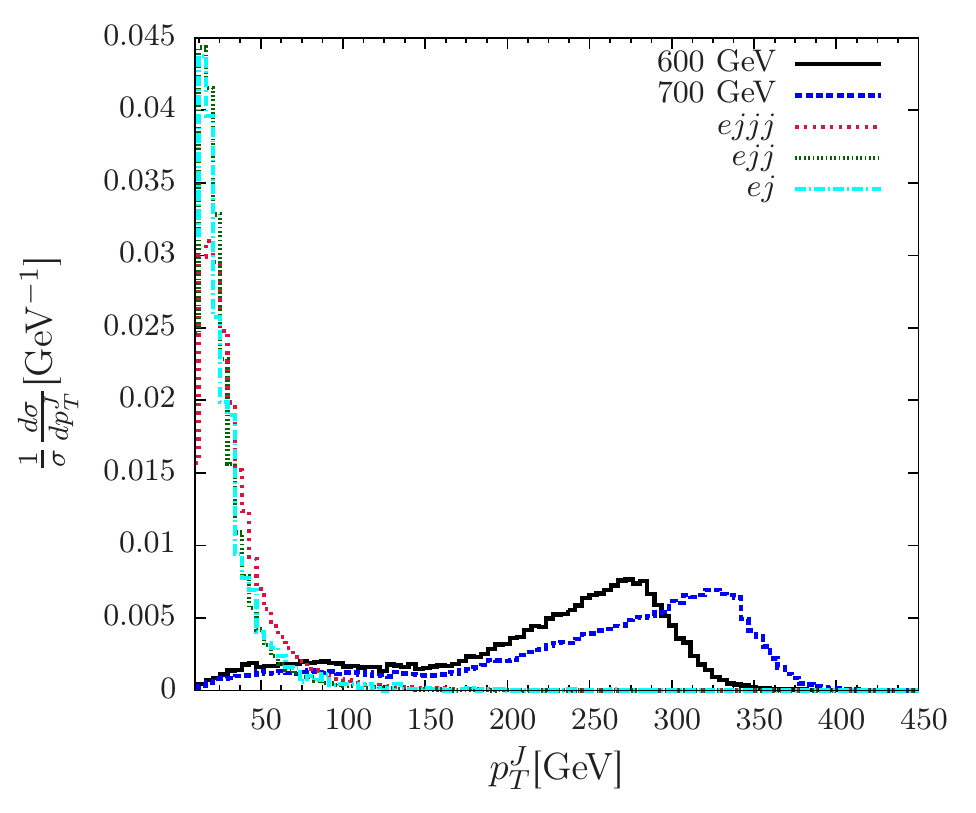}
\includegraphics[width=0.475\textwidth]{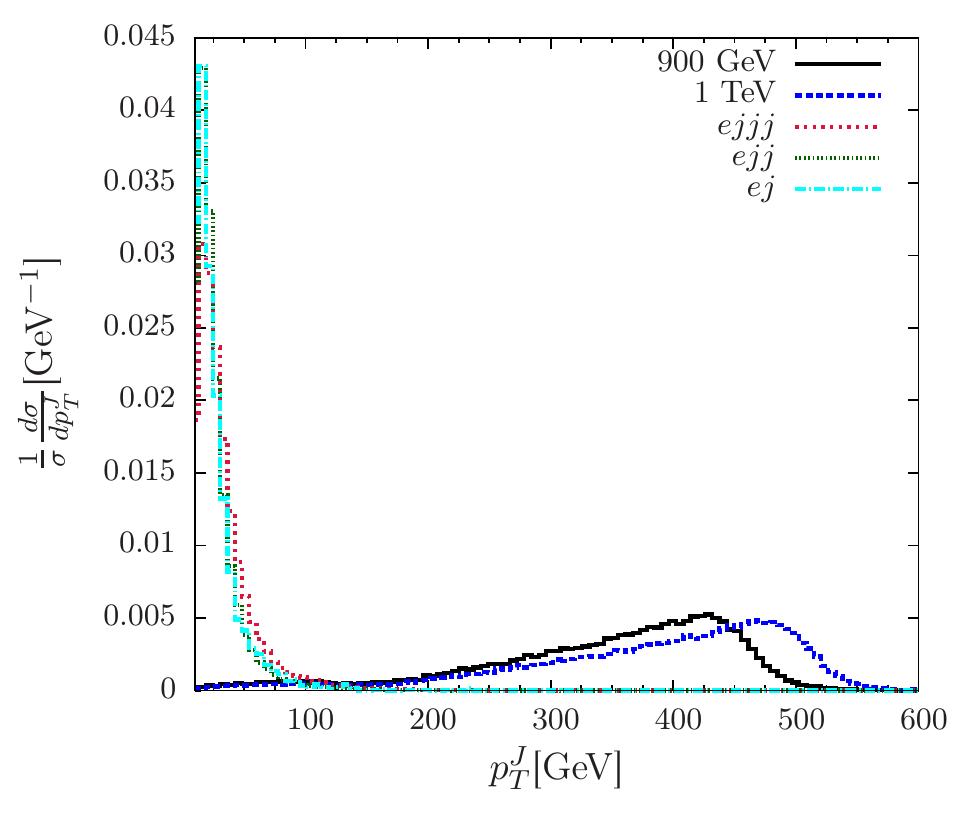}
\caption{Transverse momentum distribution of the fat jet $(p_T^J)$ from the signal and background events for $M_N=600$ GeV and $700$ GeV at the $\sqrt{s}=1.3$ TeV LHeC (left panel) 
and $M_N=900$ GeV and $1$ TeV at the $\sqrt{s}=1.8$ TeV HE-LHeC (right panel).}
\label{LHeCfig2}
\end{figure}

\begin{figure}[]
\centering
\includegraphics[width=0.475\textwidth]{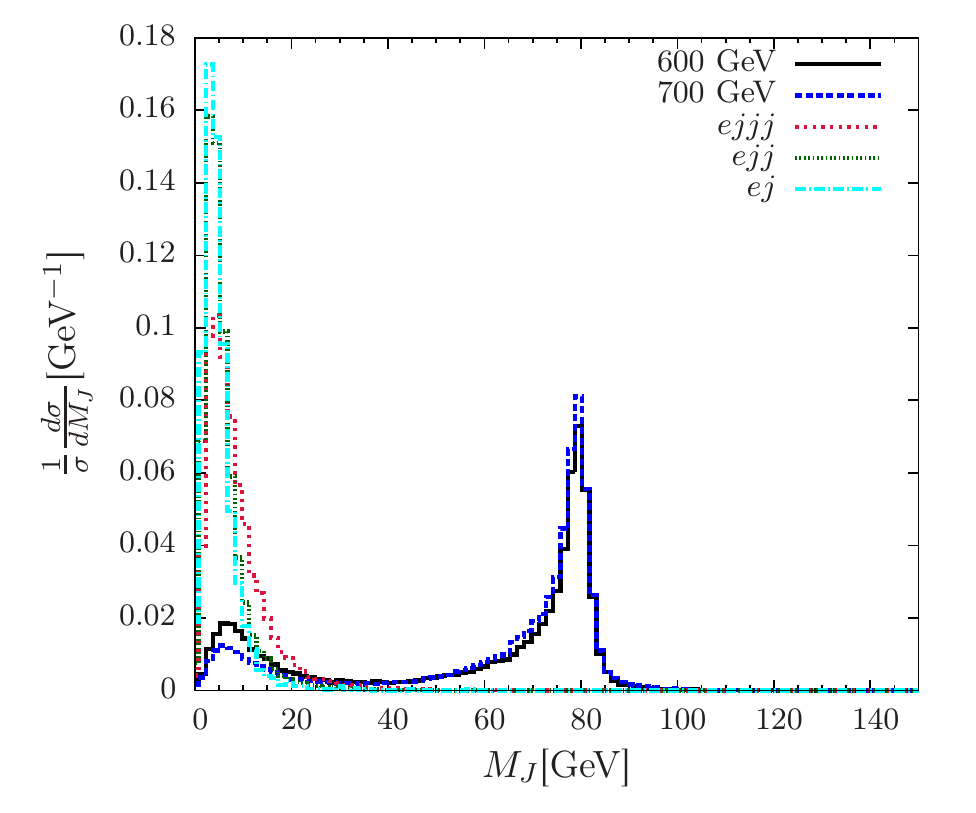}
\includegraphics[width=0.475\textwidth]{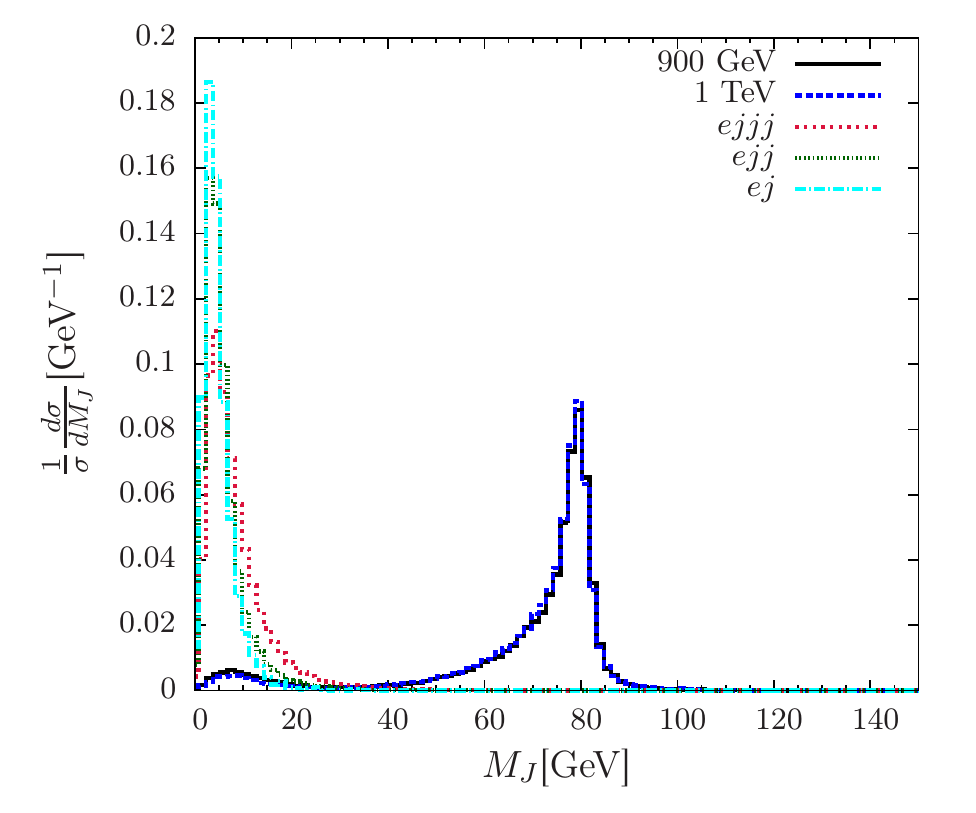}
\caption{Jet mass $(M_J)$ distribution of the fat jet from the signal and background events for $M_N=600$ GeV and $700$ GeV at the $\sqrt{s}=1.3$ TeV LHeC (left panel) and $M_N=900$ GeV and $1$ TeV at the $\sqrt{s}=1.8$ TeV HE-LHeC (right panel).}
\label{LHeCfig3}
\end{figure}

\begin{figure}[]
\centering
\includegraphics[width=0.475\textwidth]{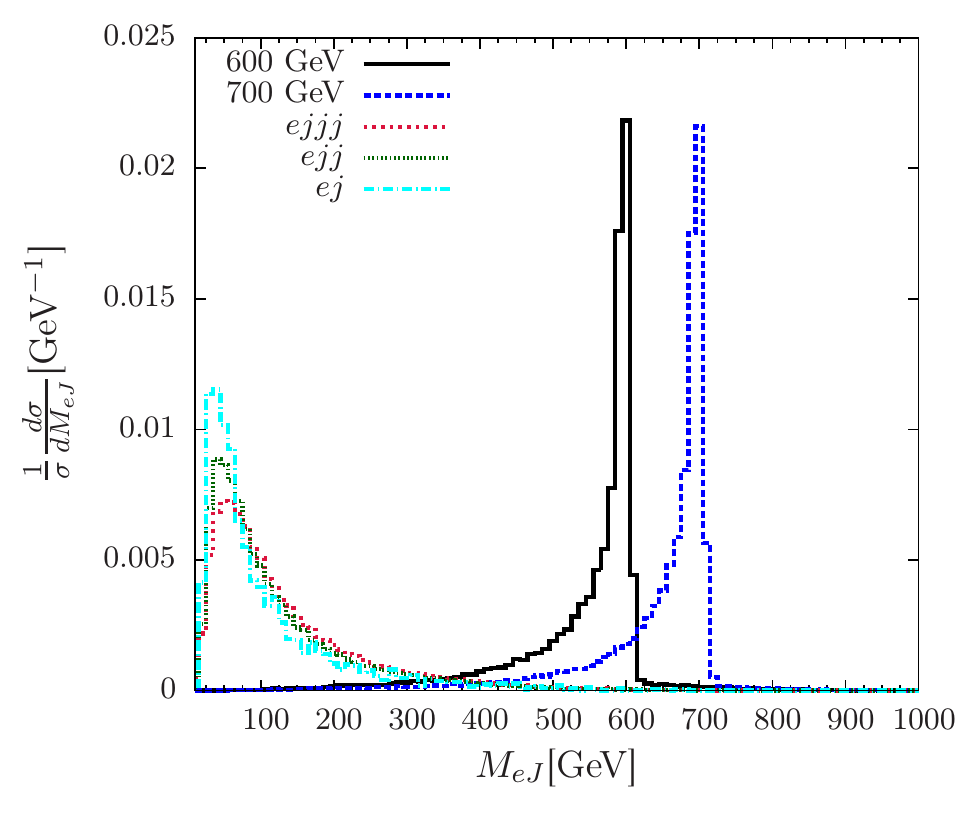}
\includegraphics[width=0.475\textwidth]{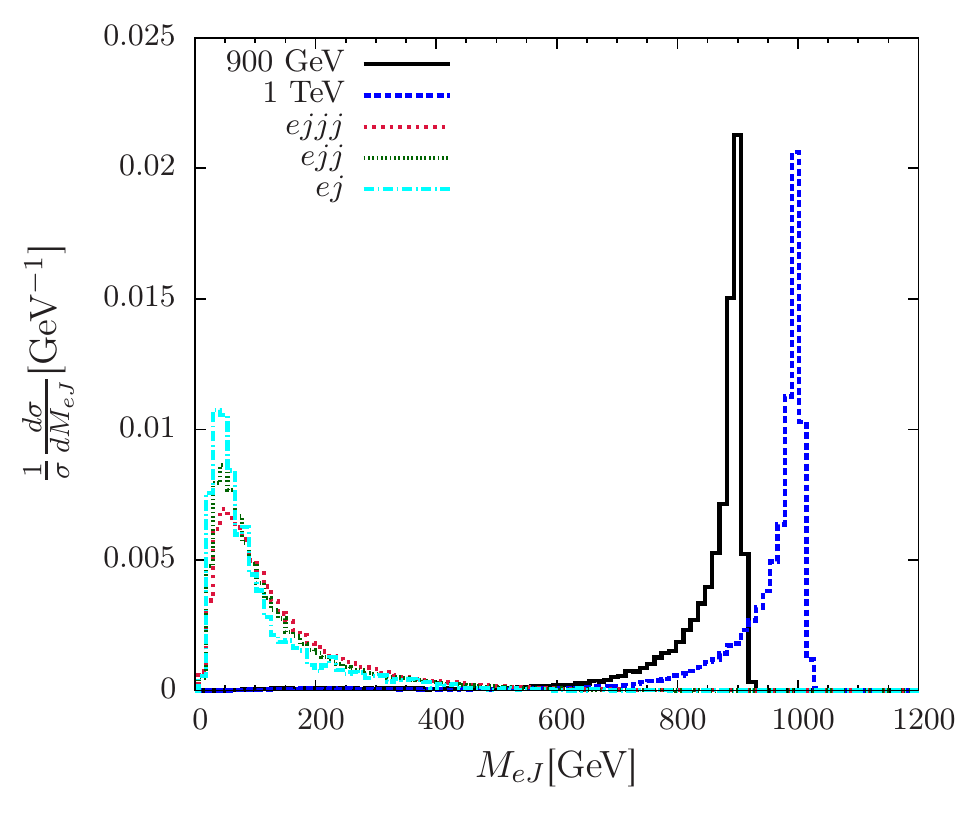}
\caption{Invariant mass distribution of the fat jet and electron system $(M_{eJ})$ from the signal and background events for $M_N=600$ GeV and $700$ GeV at the $\sqrt{s}=1.3$ TeV LHeC (left panel) and $M_N=900$ GeV and $1$ TeV at the $\sqrt{s}=1.8$ TeV HE-LHeC (right panel).}
\label{LHeCfig4}
\end{figure}

The LNV signal is almost background free until some $e^++$jets events appear from some radiations, however, that effect will be negligible. Therefore for completeness we include the LNC channel
where the leading SM backgrounds will come from $e^{-}jjj$, $e^{-}jj$ and $e^{-}j$ including initial state and final state radiations. For completeness we include both of the LNV and LNC channels.
Further we use the fat-jet algorithm to reduce the SM backgrounds. We have shown the distributions of the transverse momentum of the leading jet $(p_{T}^{j_{1}})$, lepton $(p_{T}^{e})$ and fat-jet 
$(p_{T}^{J})$ in Figs.\ref{LHeCfig0}-\ref{LHeCfig2}. The fat-jet mass distribution $(M_{J})$ has been shown in Figs.\ref{LHeCfig3}.
 The invariant mass distribution of the lepton and fat-jet system $(M_{eJ})$ has been shown in Fig.~\ref{LHeCfig4}. We have also compared the signals with the corresponding SM backgrounds.
  As a sample we consider $M_{N}=600$ GeV and $700$ GeV for $\sqrt{s}=1.3$ TeV LHeC and $M_{N}=900$ GeV, $1$ TeV at $\sqrt{s}=1.8$ TeV HE-LHeC as shown in Figs.\ref{LHeCfig0}-\ref{LHeCfig4}.

We have chosen $M_N=400$ GeV- $900$ GeV for the $1.3$ TeV LHeC and $M_N=800$ GeV- $1.5$ TeV for the $1.8$ TeV HE-LHeC. 
As benchmark points we have chosen $M_N=600$ GeV, $700$ GeV at the $1.3$ TeV LHeC and $M_N=900$ GeV, $1.0$ TeV at the $1.8$ TeV HE-LHeC after the basic cuts.
In view of the distributions in Figs.\ref{LHeCfig0}-\ref{LHeCfig4}, we have used the following advanced selection cuts to reduce the backgrounds: 
\subsubsection{Advanced cuts for $M_{N}=400$ GeV -$900$ GeV at the $\sqrt{s}=1.3$ TeV LHeC after the detector simulation}
\begin{itemize}
 \item Transverse momentum for lepton and jet, $p_{T}^{e^{\pm}}>50$ GeV.
\item Transverse momentum for fat-jet $p_{T}^{J}>175$ GeV.
 \item Fat-jet mass $M_{J}>70$ GeV.
 \item Invariant mass window of $e^{\pm}$ and fat-jet $J$, $|M_{eJ}-M_{N}|\leq 20$ GeV.
\end{itemize}
 We have noticed that $M_J > 70$ GeV cuts out the low energy peaks $(M_J \leq 25$ GeV$)$ which come from the 
hadronic activity of the low energy jets. Similarly, the $p_T^J$ and $p_T^e$ cuts are also very effective. Due to the presence of the RHN, these distributions from the signal will be in the high values than the SM backgrounds. Therefore selecting such cuts at high values, as we have done here, will be extremely useful to reduce the SM backgrounds.

We have noticed that $ej$ background can completely be reduced with the application of the kinematic cuts on $p_T^e$, $p_T^{J}$ and $M_J$. It is difficult to obtain a fat jet from this process because the $t$ channel exchange 
of the $Z$ boson and photon will contribute to this process, however, the other low-energy jets may come from the radiations at the initial and final states. These jets do not help to make the fat jets sufficiently energetic. 
Therefore $p_T^J > 175$ GeV $(p_T^J > 400$ GeV$)$ at the LHeC (HE-LHeC) are very useful. Similarly the $ejjj$ is the irreducible background in this case which will contribute most among the backgrounds.
Whereas $ejj$ is the second leading background in this case. However, both of these backgrounds can be reduced using the invariant mass cut of the RHN. As the RHN will decay according to $N \to eJ$, therefore the invariant mass 
of the $eJ$ system with an window of $20$ GeV $(|M_{eJ}-M_N| \leq 20$ GeV$)$ will be extremely useful to reduce the backgrounds further in these colliders. In Tab.~\ref{LHeC1} we have given the two benchmark scenarios at the $1.3$ TeV LHeC
where the signal events are normalized by the square of the mixing.
\begin{table*}[!htbp]
\begin{tabular}{|c|c|c|c|c|c|} 
\hline
Cuts & \multicolumn{2}{|c|}{Signal} & \multicolumn{2}{|c|}{Background} & Total \\ \hline
     &  $M_{N_{1}}=600$ GeV & $M_{N_{1}}=700$ GeV          & $ejjj$ & $ejj$ &  \\ \hline
Basic Cuts & 645,860 & 261,254   & 70,029,800 & 189,689,000 &  259,718,800 \\
 $p^{J}_{T}>175$ GeV & 476,640  &  214,520  & 295,658      &    338,720    &  634,378  \\
$M_{J}>70$ GeV & 356,350 & 160,017 & 35,244 & 17,520 & 52,764 \\
 $p_{T}^{e}>50$ GeV & 356,126 & 159,918 & 33,286 &  17,520   &  50,806     \\
$|M_{eJ}-M_{N}|\leq 20$ GeV & 304,457 & 129,690 & 7  &  1   & 8   \\\hline
\end{tabular}
\caption{Cut flow of the signal and background events for the final state $e^{\pm}+J+j_{1}$ for $M_{N}=600$ GeV and $700$ GeV with $\sqrt{s}=1.3$ TeV LHeC where the signal events are normalized by the square of the mixing.}
\label{LHeC1}
\end{table*}  
\subsubsection{Advanced cuts for $M_{N}=800\,\text{GeV}-1.5$ TeV at the $\sqrt{s}=1.8$ TeV HE-LHeC after the detector simulation}
\begin{itemize}
 \item Transverse momentum for lepton, $p_{T}^{e^{\pm}}>250$ GeV.
\item Transverse momentum for fat-jet $p_{T}^{J}>400$ GeV.
 \item Fat-jet mass $M_{J}>70$ GeV.
 \item Invariant mass window of $e^{\pm}$ and fat-jet $J$, $|M_{eJ}-M_{N}|\leq 20$ GeV.
\end{itemize}
\begin{table*}[!htbp]
\begin{tabular}{|c|c|c|c|c|c|} 
\hline
Cuts & \multicolumn{2}{|c|}{Signal} & \multicolumn{2}{|c|}{Background} & Total \\ \hline
     &  $M_{N_{1}}=900$ GeV & $M_{N_{1}}=1$ TeV          & $ejjj$ & $ejj$ &  \\ \hline
 Basic Cuts & 427,311 & 207,015   & 108,243,000 & 273,410,000 & 381,653,000 \\
 $p^{J}_{T}>400$ GeV & 158,694   &  110,289  &  12,225   &  12,450   &  24,675    \\
$M_{J}>70$ GeV & 145,558 & 96,787 & 4,596 & 4,150 & 8,746 \\
$p_{T}^{e}>250$ GeV & 144,997 & 96,487 & 4,596 &  4,150   &  8,746      \\
$|M_{eJ}-M_{N}|\leq 20$ GeV & 119,659 & 71,490 & 3  &  1   & 4   \\
\hline
\end{tabular}
\caption{Cut flow of the signal and background events for the final state $e^{\pm}+J+j_{1}$ for $M_{N}=900$ GeV and $1.0$ TeV with $\sqrt{s}=1.8$ TeV HE-LHeC where the signal events are normalized by the square of the mixing.}
\label{LHeC2}
\end{table*}  

We have chosen $M_{N}=900~\text{GeV} \text{and}\,1$ TeV at the $\sqrt{s}=1.8$ TeV HE-LHeC. The corresponding signals normalized by the square of the mixing 
and the SM backgrounds are listed in Tab.~\ref{LHeC2}. Due to the heavier mass range of the RHN, we have chosen stronger cuts for the transverse momenta of the electron and fat-jet which became useful to reduce the backgrounds.

\subsection{Linear collider analysis for the signal $e^{\pm}+J+p_{T}^{miss}$}

\begin{figure}[]
\centering
\includegraphics[width=0.475\textwidth]{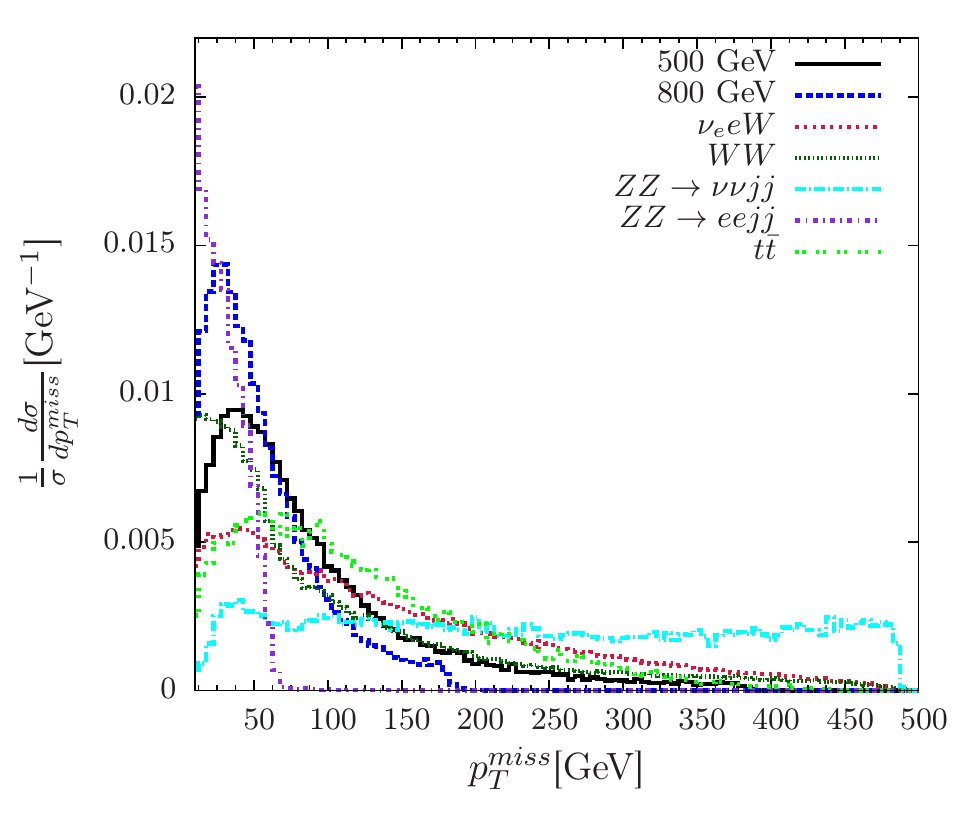}
\includegraphics[width=0.475\textwidth]{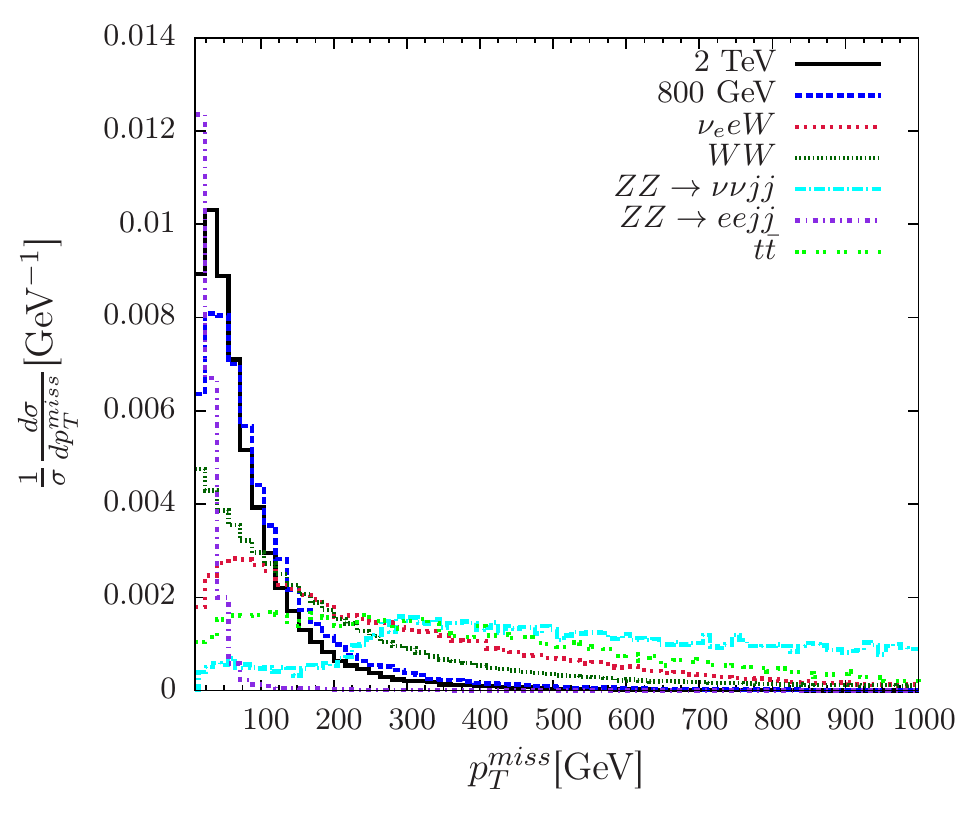}
\caption{Missing momentum distribution of the signal and background events for $M_N=500$ GeV and $800$ GeV at the $\sqrt{s}=1$ TeV (left panel) and $M_N=800$ GeV and $2$ TeV at the $\sqrt{s}=3$ TeV (right panel) linear colliders.}
\label{ILC1}
\end{figure}

\begin{figure}[]
\centering
\includegraphics[width=0.475\textwidth]{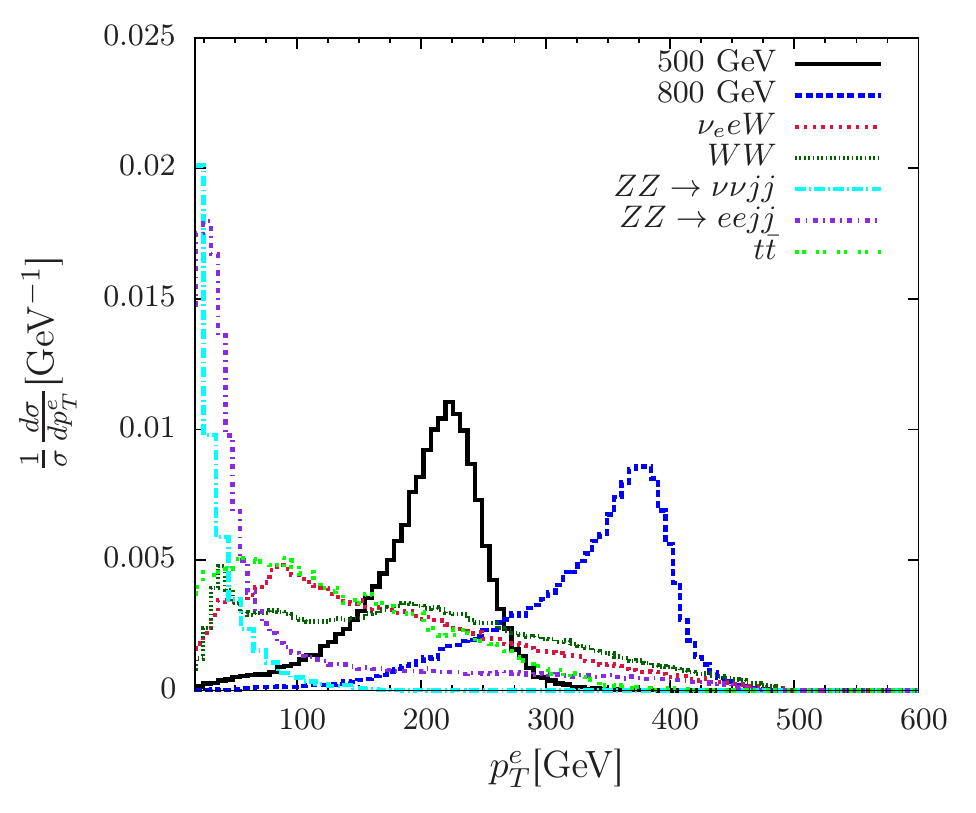}
\includegraphics[width=0.475\textwidth]{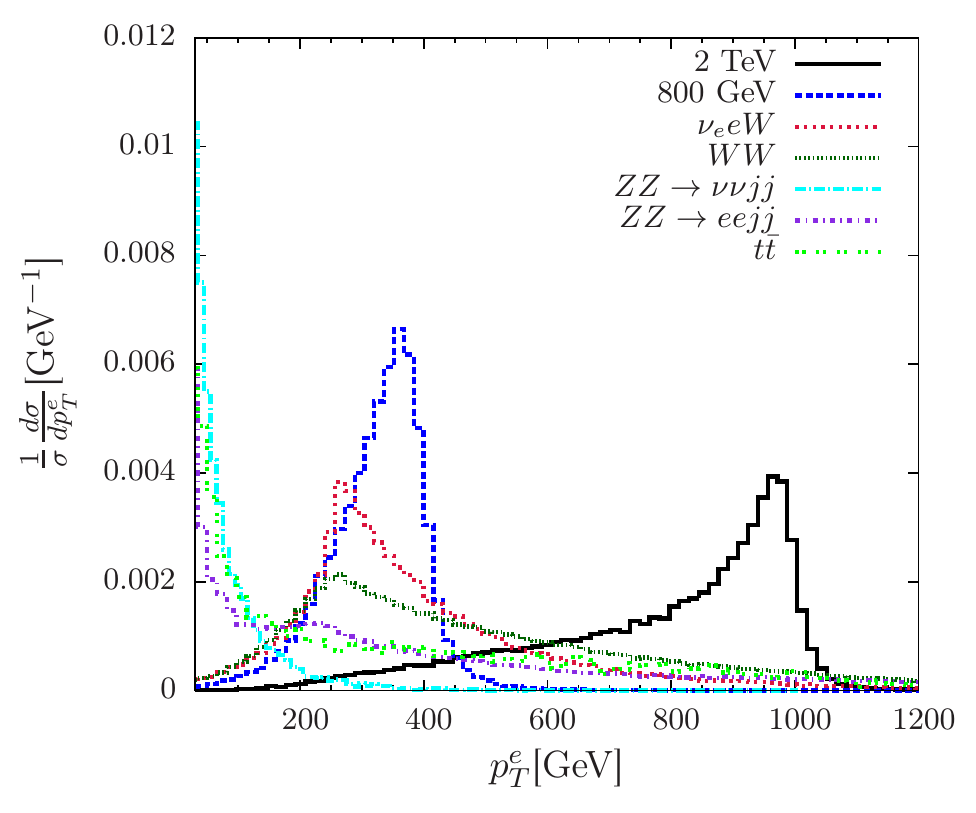}
\caption{Transverse momentum distribution of the electron $(p_T^e)$ from the signal and background events for $M_N=500$ GeV and $800$ GeV at the $\sqrt{s}=1$ TeV (left panel) and $M_N=800$ GeV and $2$ TeV at the $\sqrt{s}=3$ TeV (right panel) linear colliders.}
\label{ILC2}
\end{figure}

\begin{figure}[]
\centering
\includegraphics[width=0.475\textwidth]{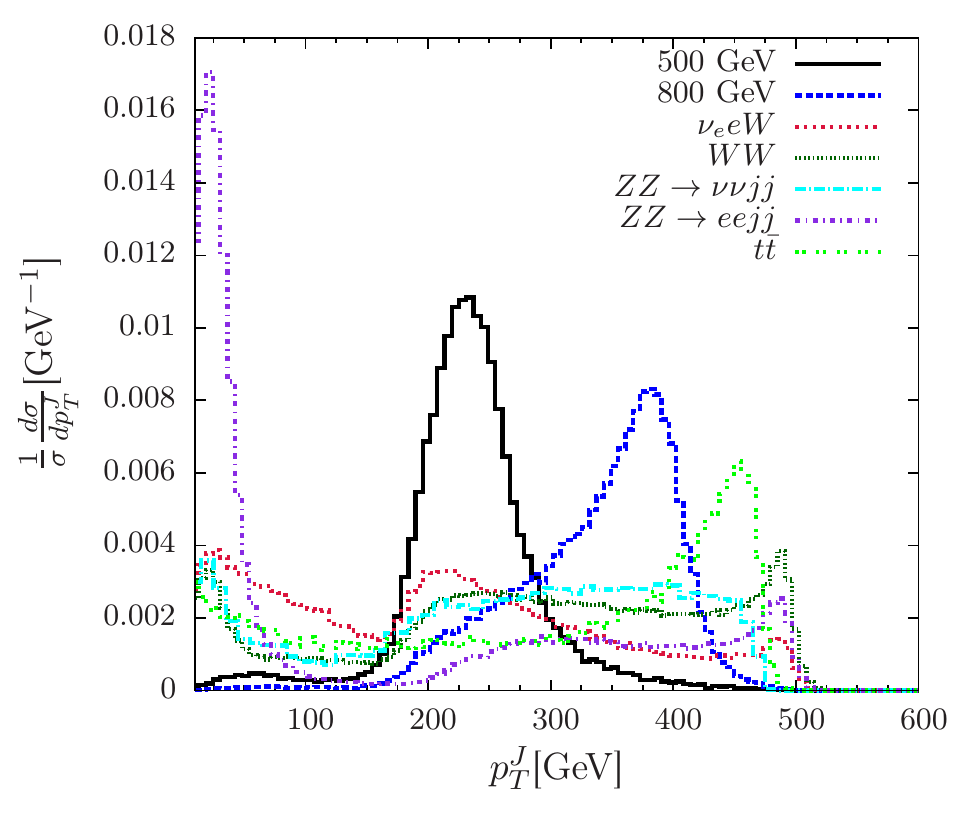}
\includegraphics[width=0.475\textwidth]{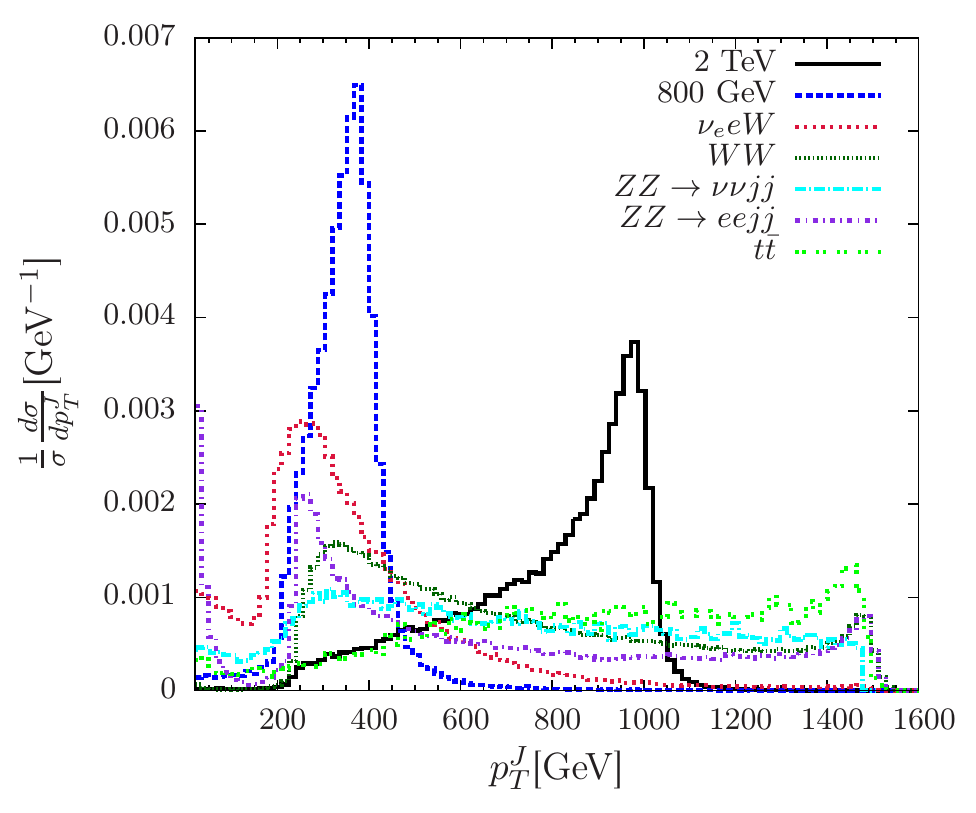}
\caption{Transverse momentum distribution of the fat jet $(p_T^J)$ from the signal and background events for $M_N=500$ GeV and $800$ GeV at the $\sqrt{s}=1$ TeV (left panel) and $M_N=800$ GeV and $2$ TeV at the $\sqrt{s}=3$ TeV linear colliders.}
\label{ILC3}
\end{figure}

\begin{figure}[]
\centering
\includegraphics[width=0.475\textwidth]{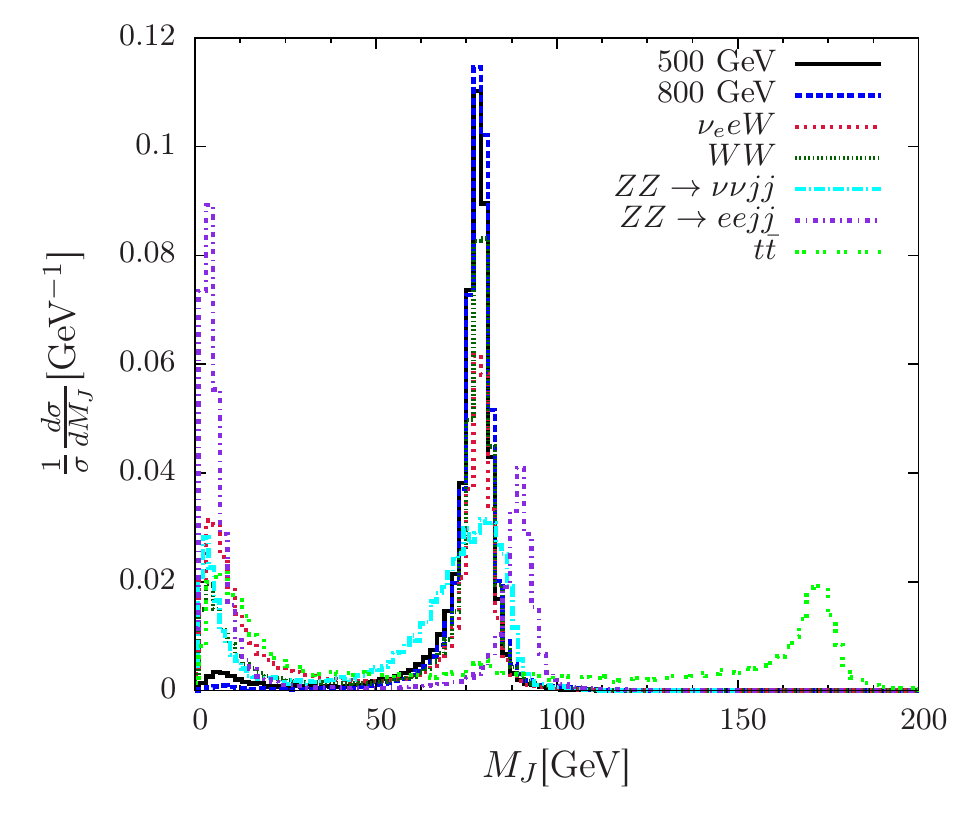}
\includegraphics[width=0.475\textwidth]{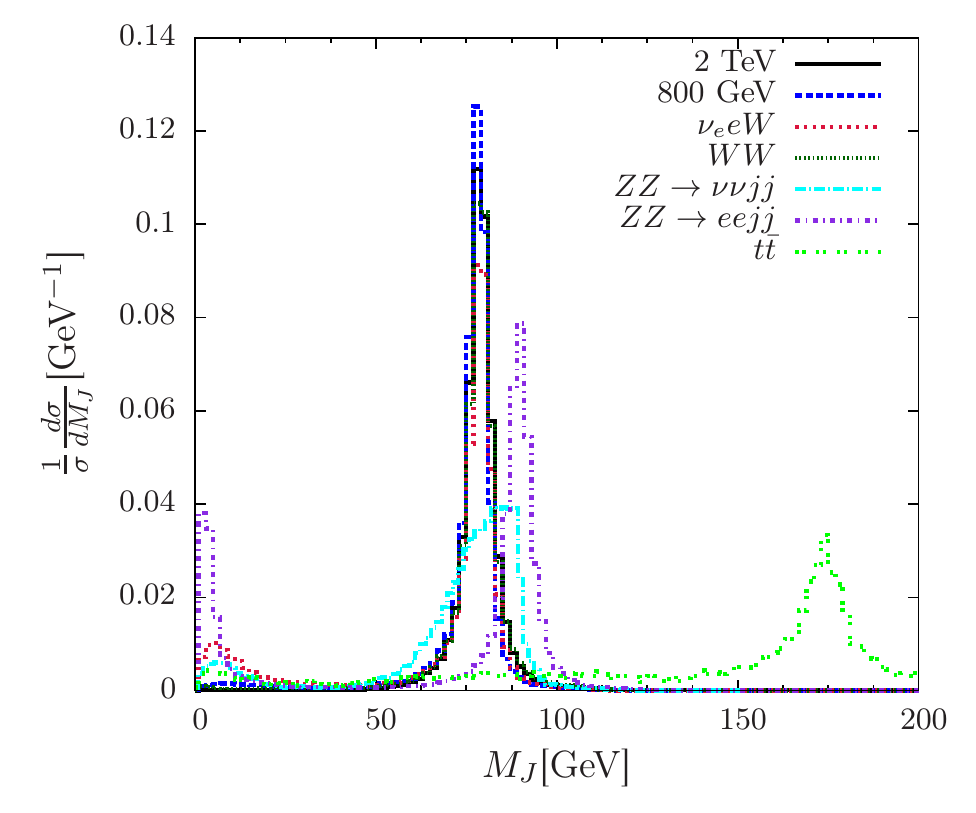}
\caption{Jet mass $(M_J)$ distribution of the fat jet from the signal and background events for $M_N=500$ GeV and $800$ GeV at the $\sqrt{s}=1$ TeV (left panel) and $M_N=800$ GeV and $2$ TeV at the $\sqrt{s}=3$ TeV (right panel) linear colliders.}
\label{ILC4}
\end{figure}

\begin{figure}[]
\centering
\includegraphics[width=0.475\textwidth]{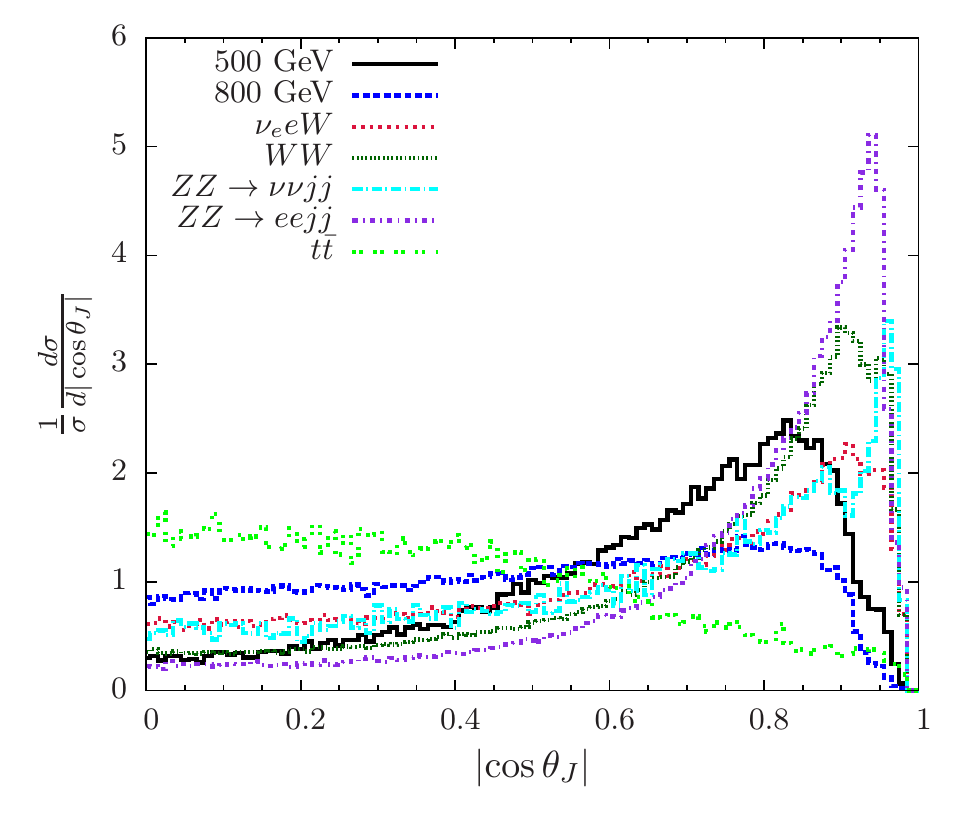}
\includegraphics[width=0.475\textwidth]{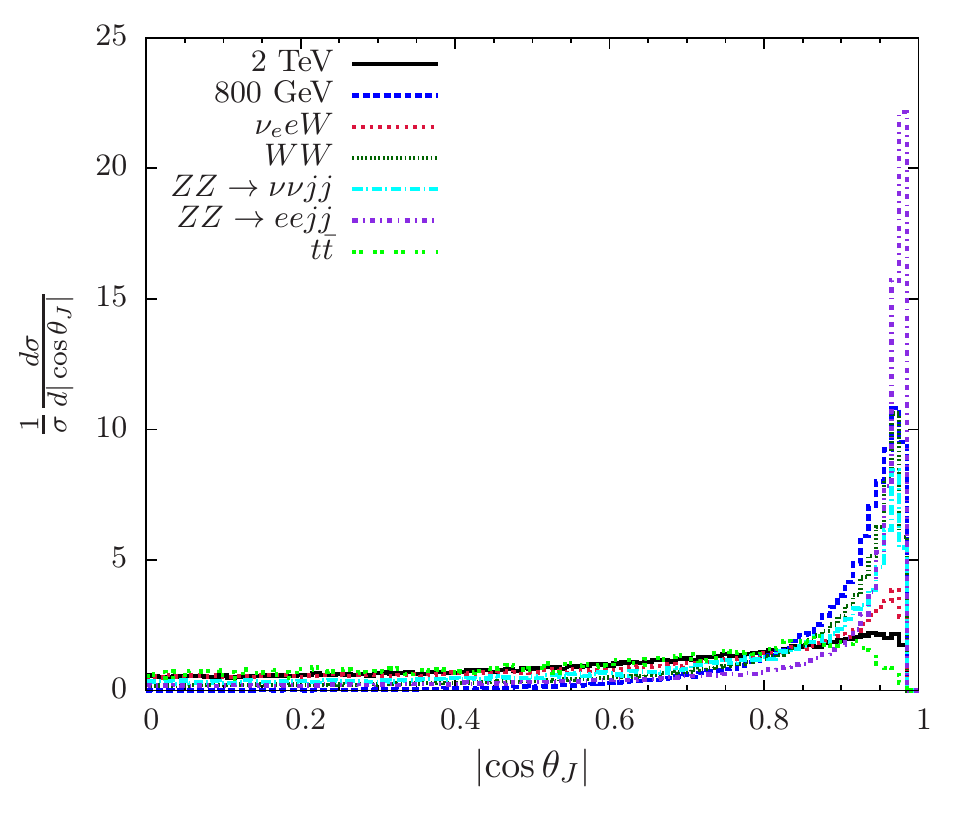}\\
\includegraphics[width=0.475\textwidth]{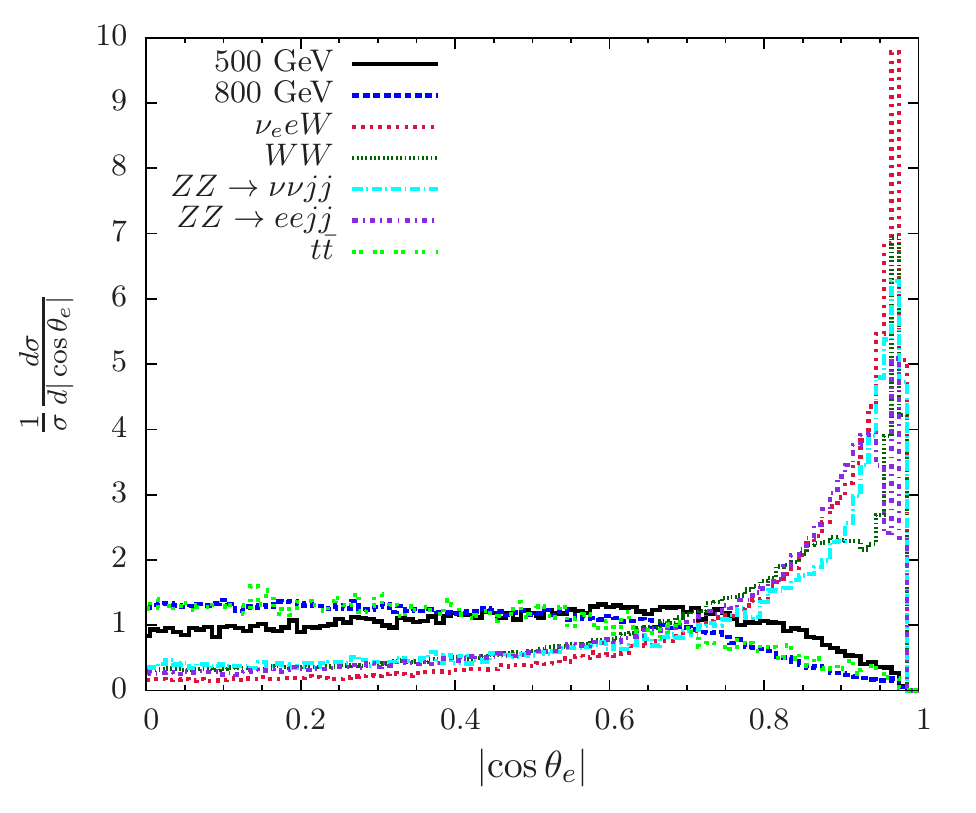}
\includegraphics[width=0.475\textwidth]{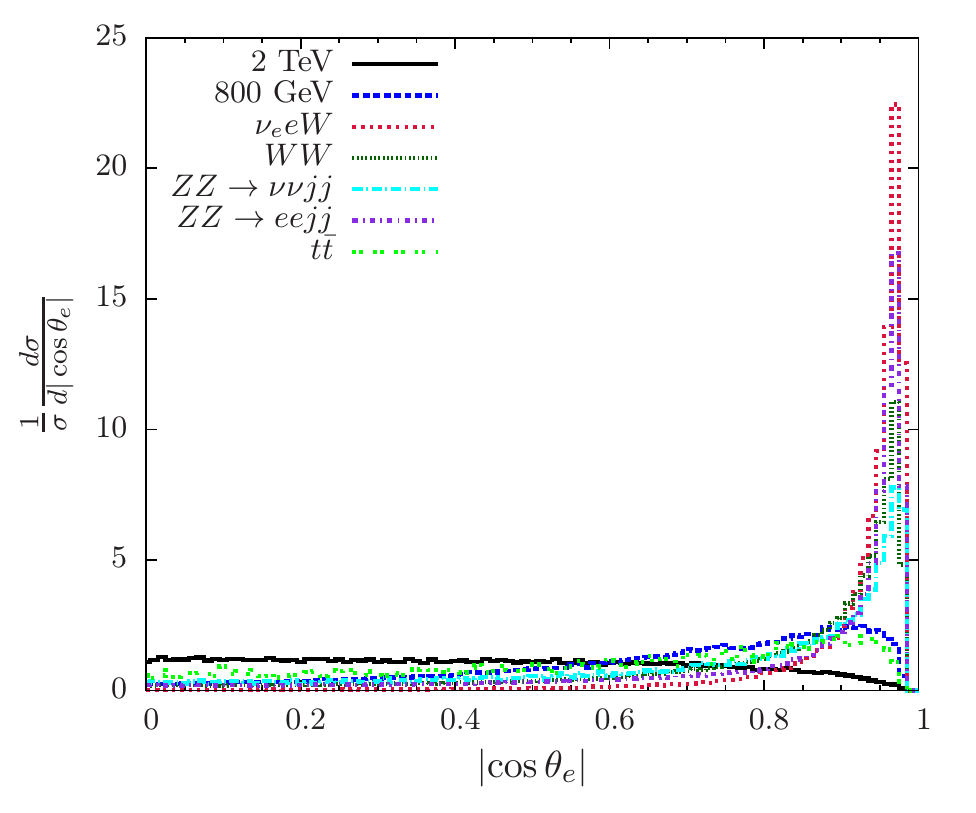}
\caption{$\cos\theta_{J(e)}$ distributions for the $J (e)$ in the first row (second row) for the $1$ TeV (left column) and $3$ TeV (right column) linear colliders.}
\label{ILC5}
\end{figure}

In linear collider we study $e^\pm+J+p_T^{miss}$ signal from the leading decay mode of the RHN at the $1$ TeV and $3$ TeV center of mass energy. 
The corresponding distributions for two benchmark points for $M_{N}=500$ GeV, $800$ GeV at$\sqrt{s}=1$ TeV and $M_{N}=800$ GeV, $2$ TeV at $\sqrt{s}=3$ TeV linear colliders are given in Figs.\ref{ILC1}-\ref{ILC5} after the basic cuts.
We perform a complete cut based analysis for the signal and the SM backgrounds. In this process we have $\nu_{e}eW$ as the leading background where as $WW$, $ZZ$ and $t\bar{t}$ are other important backgrounds.

We have shown the missing momentum $(p_T^{miss})$, transverse momenta of the electron $p_{T}^{e}$ and fat-jet $p_{T}^{J}$ in Figs.~\ref{ILC1}-\ref{ILC3} for the linear colliders.
The fat-jet mass $M_{J}$ distribution has been shown in Fig.~\ref{ILC4}. We construct the polar angle variable in Fig.~\ref{ILC5} for the electron (fat jet), $\cos\theta_e (\cos\theta_J)$ where $\theta_{e(J)}=\tan^{-1}\Big[\frac{p_T^{e(J)}}{p_z^{e(J)}}\Big]$,
where $p_z^{e(J)}$ is the $z$ component of the three momentum of the electron (fat jet). This is a very effective cut which reduces the SM background significantly.
In view of these distributions, we have used the following advanced selection cuts to reduce the backgrounds:
\subsubsection{Advanced cuts for $M_{N}=400$ GeV-$900$ GeV at the $\sqrt{s}=1$ TeV linear collider after the detector simulation}
\begin{itemize}
\item Transverse momentum for fat-jet $p_{T}^{J}>150$ GeV for $M_{N}$ mass range $400$ GeV-$600$ GeV and $p_{T}^{J}>250$ GeV for $M_{N}$ mass range $700$ GeV-$900$ GeV.
 \item Transverse momentum for leading lepton $p_{T}^{e^{\pm}}>100$ GeV for $M_{N}$ mass range $400$ GeV-$600$ GeV and $p_{T}^{e^{\pm}}>200$ GeV for $M_{N}$ mass range $700$ GeV-$900$ GeV.
 \item Polar angle of lepton and fat-jet $|\text{cos}~\theta_{e}|<0.85$, $|\text{cos}~\theta_{J}|<0.85$.
 \item Fat-jet mass $M_{J}>70$ GeV.
\end{itemize}
We have tested $M_N=400$ GeV to $900$ GeV at the $\sqrt{s}=1$ TeV at the linear collider.
Hence we consider two benchmark points at the $\sqrt{s}=1$ TeV linear collider such as $M_N=500$ GeV and $800$ GeV.
The cut flow for the $\sqrt{s}=1$ TeV are given in the Tabs.~\ref{ILC5001TeV} and \ref{ILC8001TeV} respectively.
We have noticed that $\cos\theta_{e(J)}$ is a very important kinematic variable and setting $|\cos\theta_{e(J)}| < 0.85$ puts a very strong cut for the SM backgrounds. 
The $M_J > 70$ GeV is also effective to cut out the low mass peaks $(1$ GeV $\leq M_J \leq 25$ GeV $)$ from the low energy jets.

\begin{table*}[!htbp]
\begin{tabular}{|c|c|c|c|c|c|c|} 
\hline
Cuts & Signal & \multicolumn{4}{|c|}{Background} & Total \\ \hline
     &            & $\nu_{e}eW$ & $WW$ & $ZZ$ & $t\bar{t}$ &  \\ \hline
Basic Cuts & 12,996,200  & 201,586 & 72,244 & 7,200 & 4,300 & 285,330 \\
$|\text{cos}~\theta_{J}|\leq 0.85$ &  12,789,800   &  148,802   & 44,910     & 3,800   & 4,100        &    201,600     \\
$|\text{cos}~\theta_{e}|\leq\,0.85$ & 12,671,800 & 79,008 & 40,574 & 2,800 & 3,900 & 126,280 \\ 
$p^{J}_{T}>150$ GeV & 12,308,300   &  70,669  &  40,490  &  2,300   &   3,200   &  116,660  \\
$M_{J}>70$ GeV & 10,923,100  & 62,303   & 37,043     & 2,100    & 2,300 &  103,700\\
$p_{T}^{\ell}>100\,\text{GeV}$ & 10,714,500       & 57,076   & 33,488     & 1,400     & 1,530  &  93,400\\ 
\hline
\end{tabular}
\caption{Cut flow for the signal and background events for the final state $e^{\pm}+J+p_T^{miss}$ for $M_{N}=500$ GeV at the $\sqrt{s}=1$ TeV linear collider. The signal events are normalized by the square of the mixing.}
\label{ILC5001TeV}
\end{table*} 
\begin{table*}[!htbp]
\begin{tabular}{|c|c|c|c|c|c|c|} 
\hline
Cuts & Signal & \multicolumn{4}{|c|}{Background} & Total \\ \hline
     &            & $\nu_{e}eW$ & $WW$ & $ZZ$ & $t\bar{t}$ &  \\ \hline
Basic Cuts & 8,684,990   &  201,586 & 72,244 & 7,200 & 4,300 & 285,330 \\
$|\text{cos}~\theta_{J}|\leq 0.85$ &  8,649,570   &  148,802   & 44,910     & 3,800   & 4,100        &    201,600     \\
$|\text{cos}~\theta_{e}|\leq\,0.85$ & 8,618,420  & 79,008 & 40,574 & 2,800 & 3,900 & 126,280 \\ 
$p^{J}_{T}>250$ GeV & 7,681,440   &  59,001  &  40,329  &  2,303   &   2,720   & 104,354   \\
$M_{J}>70$ GeV & 7,176,280  & 53,990   & 36,997      & 2,187    & 2,282 & 95,437 \\
$p_{T}^{\ell}>200\,\text{GeV}$ & 7,080,200       & 38,729   & 26,208      & 942     & 613  & 66,493 \\  
\hline
\end{tabular}
\caption{Cut flow for the signal and background events for the final state $e^{\pm}+J+p_T^{miss}$ for $M_{N}=800$ GeV at the $\sqrt{s}=1$ TeV linear collider. The signal events are normalized by the square of the mixing.}
\label{ILC8001TeV}
\end{table*}  
\subsubsection{Advanced cuts for $M_{N}=700$ GeV-$2.9$ TeV at the $\sqrt{s}=3$ TeV linear collider after the detector simulation}
\begin{itemize}
\item Transverse momentum for fat-jet $p_{T}^{J}>250$ GeV for the $M_{N}$ mass range $700$ GeV-$900$ GeV and $p_{T}^{J}>400$ GeV for $M_{N}$ mass range $1-2.9$ TeV.
 \item Transverse momentum for leading lepton $p_{T}^{e^{\pm}}>200$ GeV for $M_{N}$ mass range $700-900$ GeV and $p_{T}^{e^{\pm}}>250$ GeV for $M_{N}$ mass range $1-2.9$ TeV.
 \item Polar angle of lepton and fat-jet $|\text{cos}~\theta_{e}|<0.85$, $|\text{cos}~\theta_{J}|<0.85$.
 \item Fat-jet mass $M_{J}>70$ GeV.
\end{itemize}
We have tested $M_N=700$ GeV to $2.9$ TeV at the $\sqrt{s}=3$ TeV at the linear collider.
Hence we consider two benchmark points at the $\sqrt{s}=3$ TeV linear collider such as $M_N=800$ GeV and $2$ TeV.
The cut flow for the benchmark points at the $\sqrt{s}=3$ TeV are given in the Tabs.~\ref{ILC8003TeV} and \ref{ILC20003TeV} respectively.
At the $3$ TeV we see almost the same behavior for the kinematic variables as we noticed at the $1$ TeV case except the $p_T$ distributions of the electron and fat jet.
\begin{table*}[!htbp]
	\begin{tabular}{|c|c|c|c|c|c|c|} 
		\hline
		Cuts & Signal & \multicolumn{4}{|c|}{Background} & Total \\ \hline
		&            & $\nu_{e}eW$ & $WW$ & $ZZ$ & $t\bar{t}$ &  \\ \hline
		Basic Cuts & 21,789,900   & 193,533 & 12,135  & 1,361     & 271 & 207,301 \\
		$|\text{cos}~\theta_{J}|\leq 0.85$   & 13,599,300      &  126,980     &  4,766     &  406    &  215     & 132,367       \\
		$|\text{cos}~\theta_{e}|\leq\,0.85$ & 12,163,300  & 21,110 & 4,609 & 390 & 195 & 26,304 \\ 
		 $p^{J}_{T}>250$ GeV &  12,083,500 &  18,619 & 4,607 & 390 & 189 & 23,807 \\
		$M_{J}>70$ GeV& 11,287,000  & 17,442   & 4,411     & 385    & 176 & 22,416 \\
		$p_{T}^{\ell}>200\,\text{GeV}$ & 11,094,300       & 16,915   & 4,108      & 343  & 104  & 21,470 \\  				
		\hline
	\end{tabular}
\caption{Cut flow for the signal and background events for the final state $e^{\pm}+J+p_T^{miss}$ for $M_{N}=800$ GeV at the $\sqrt{s}=3$ TeV linear collider. The signal events are normalized by the square of the mixing. }
\label{ILC8003TeV}
\end{table*}  
\begin{table*}[!htbp]
\begin{tabular}{|c|c|c|c|c|c|c|} 
	\hline
	Cuts & Signal & \multicolumn{4}{|c|}{Background} & Total \\ \hline
	&            & $\nu_{e}eW$ & $WW$ & $ZZ$ & $t\bar{t}$ &  \\ \hline
	Basic Cuts & 13,822,500   & 193,533 & 12,135  & 1,382     & 271 & 207,322 \\
	$|\text{cos}~\theta_{J}|\leq 0.85$  &  12,701,600       &  126,980     &  4,766     &  412    & 215     &  132,374   \\
	$|\text{cos}~\theta_{e}|\leq\,0.85$ & 12,647,200  & 21,110 & 4,609 & 396 & 195 & 26,310 \\ 
	$p^{J}_{T}>400$ GeV & 12,611,000  & 15,737 & 4,605 & 396 & 184 & 20,923 \\
	$M_{J}>70$ GeV & 12,015,600  & 14,889   & 4,410      & 391  & 175 & 19,865 \\
	$p_{T}^{\ell}>250\,\text{GeV}$ & 11,987,000       & 14,184   & 4,010      & 336    & 10  & 18,630 \\		
	\hline
\end{tabular}
\caption{Cut flow for the signal and background events for the final state $e^{\pm}+J+p_T^{miss}$ for $M_{N}=2$ TeV at the $\sqrt{s}=3$ TeV linear collider. The signal events are normalized by the square of the mixing.}
\label{ILC20003TeV}
\end{table*}  
At this point we must mention that the backgrounds like $ZZ$ and $t\bar{t}$ can have more than one lepton in the final state which has been efficiently vetoed to reduce the effect.
\subsection{Linear collider analysis for the signal $J_{b}+p_{T}^{miss}$}
Considering the $N\to h\nu, h \to J_b$ mode at the linear collider we obtain the $J_{b}+p_{T}^{miss}$ final state.
For this final state the dominant SM backgrounds come from the processes $h\nu_{\ell}\bar{\nu}_{\ell}$ and $Z\nu_{\ell}\bar{\nu}_{\ell}$. 
Backgrounds can also come from the intermediate processes $ZZ$ and $ZH$. 
We have generated the background events combining all these processes in MadGraph for our analysis.
\begin{figure}[h]
\centering
\includegraphics[width=0.475\textwidth]{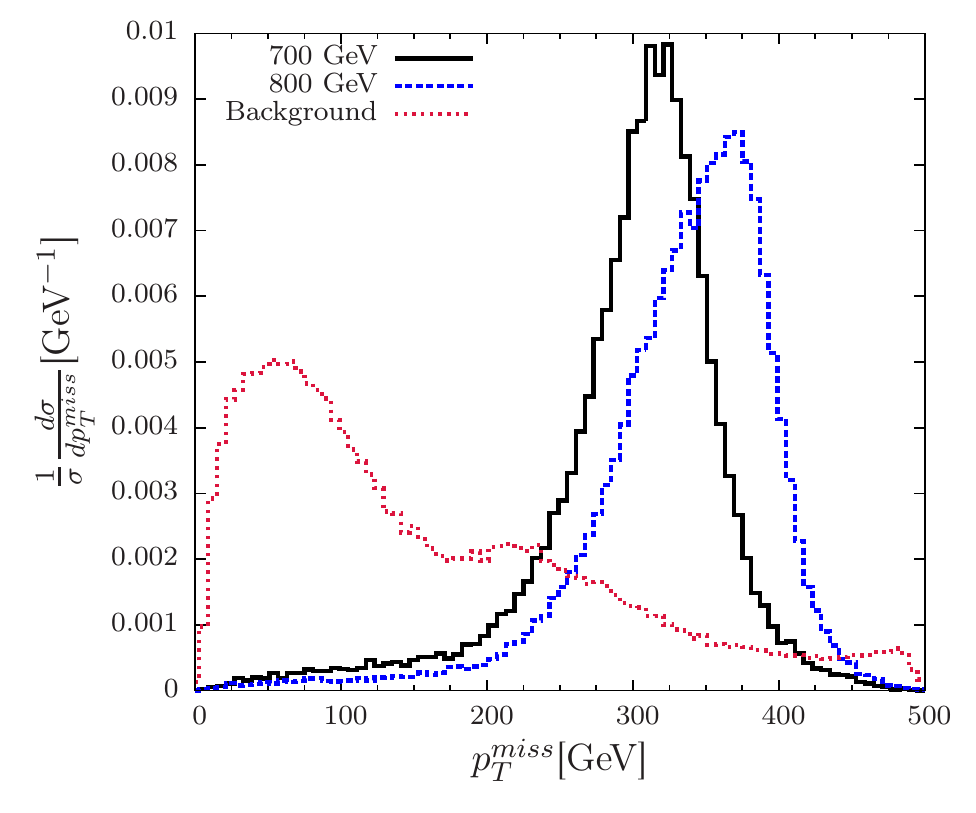}
\includegraphics[width=0.475\textwidth]{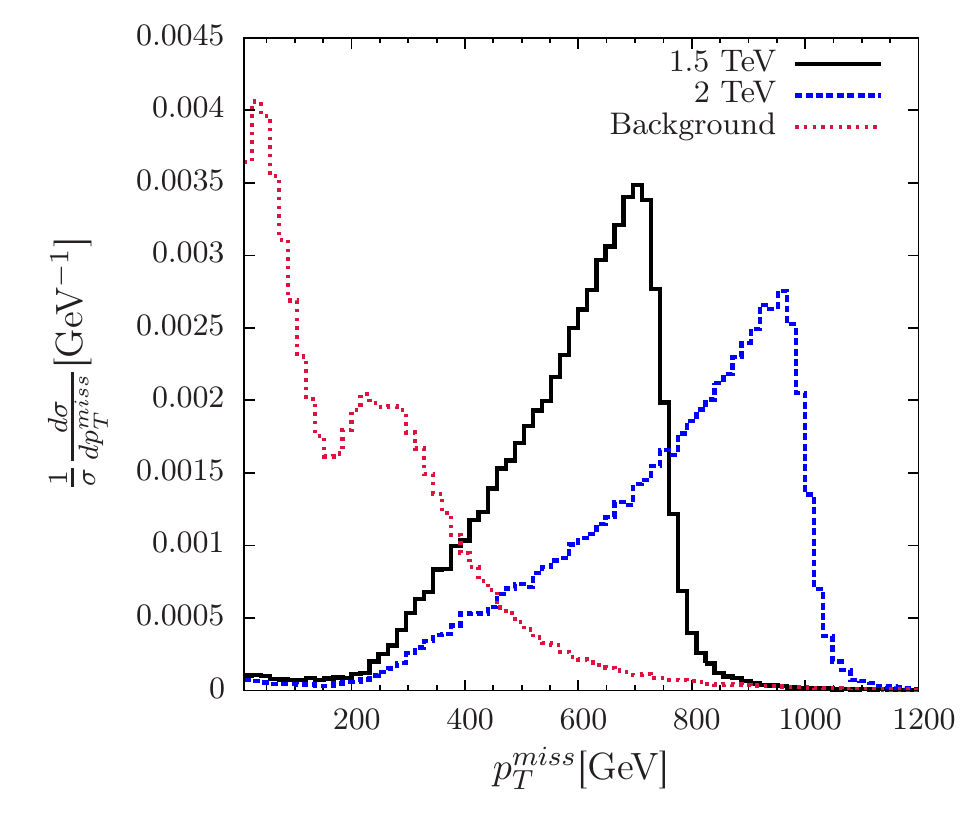}
\caption{$p_T^{miss}$ distribution of the signal and background events for $M_N=700$ GeV and $800$ GeV at the $\sqrt{s}=1$ TeV (left panel) and $M_N=1.5$ TeV and $2$ TeV at the $\sqrt{s}=3$ TeV (right panel) linear colliders.}
\label{ILC METfatb histogram}
\end{figure}
\begin{figure}[h]
\centering
\includegraphics[width=0.475\textwidth]{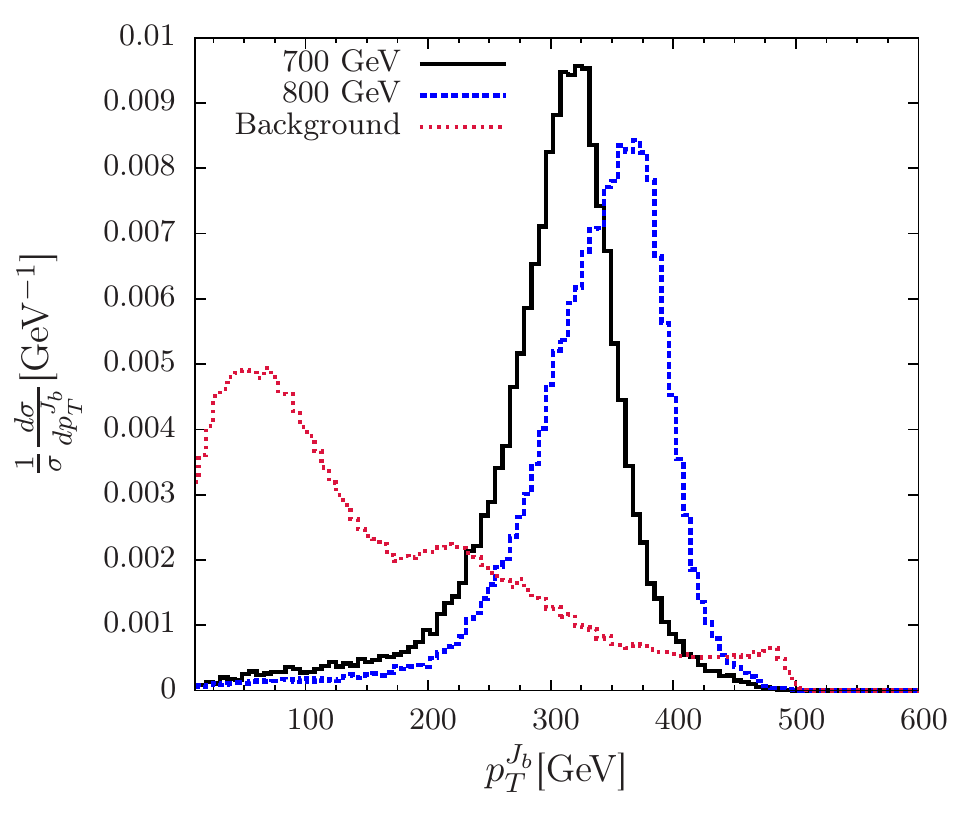}
\includegraphics[width=0.475\textwidth]{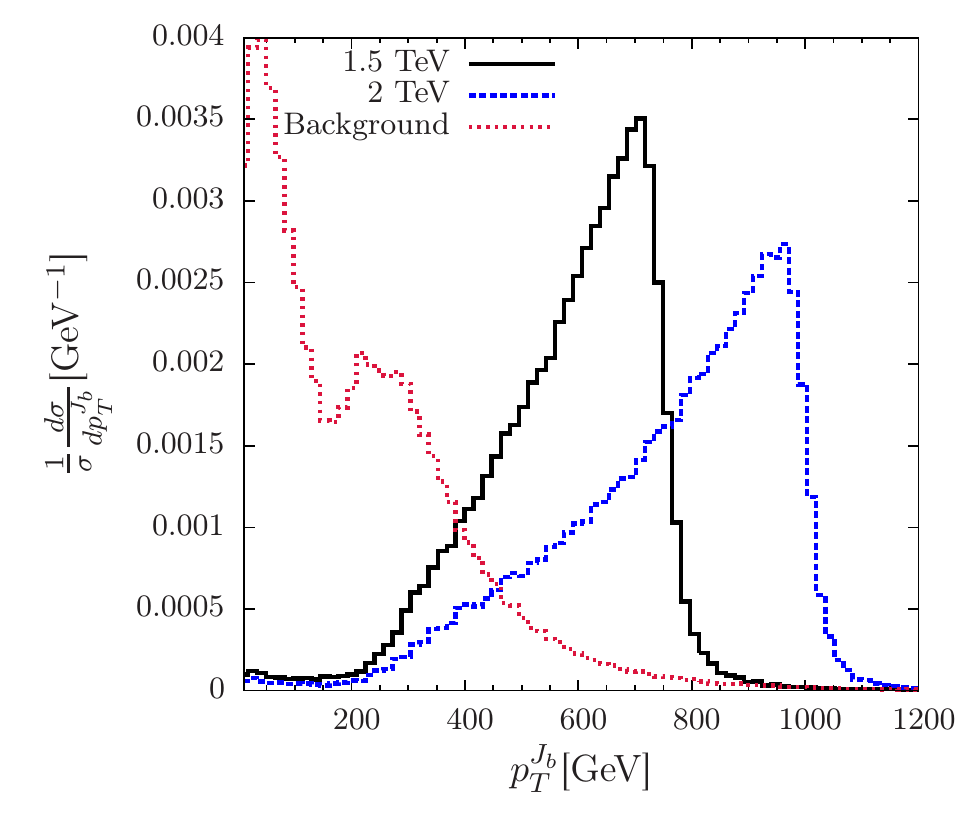}
\caption{Transverse momentum distribution of $J_b$ $(p_{T}^{J_{b}})$ from the signal and background events for $M_N=700$ GeV and $800$ GeV at the $\sqrt{s}=1$ TeV (left panel) and $M_N=1.5$ TeV and $2$ TeV at the $\sqrt{s}=3$ TeV (right panel) linear colliders.}
\label{ILC ptfatb histogram}
\end{figure}
\begin{figure}[h]
\centering
\includegraphics[width=0.475\textwidth]{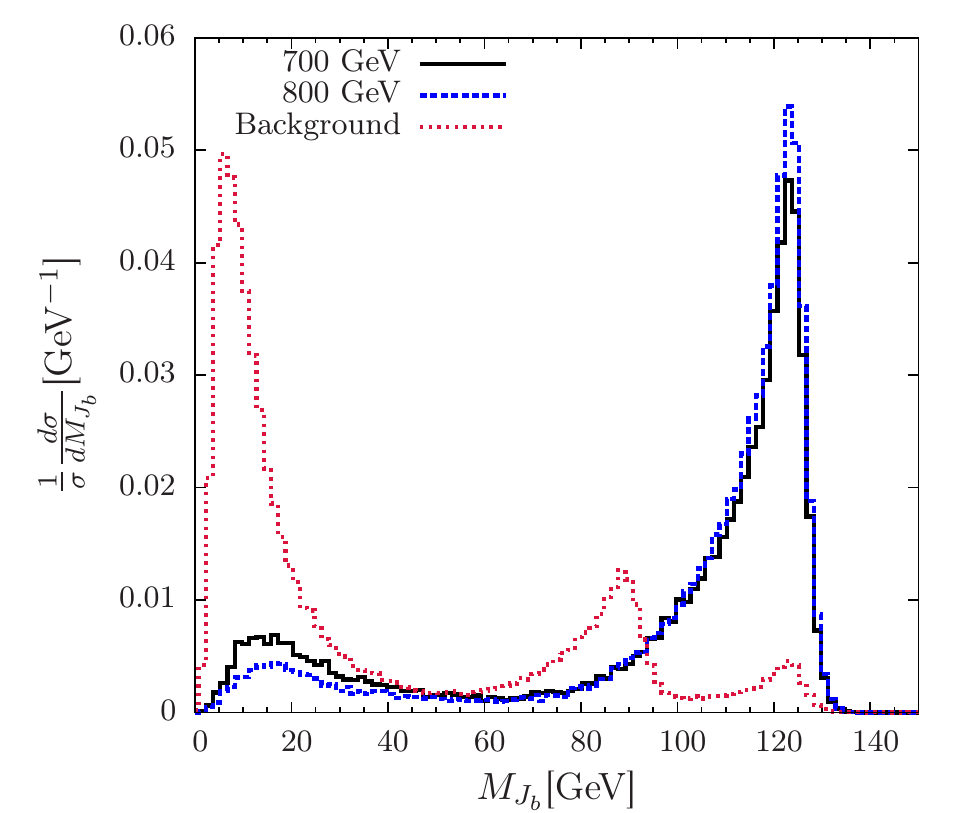}
\includegraphics[width=0.475\textwidth]{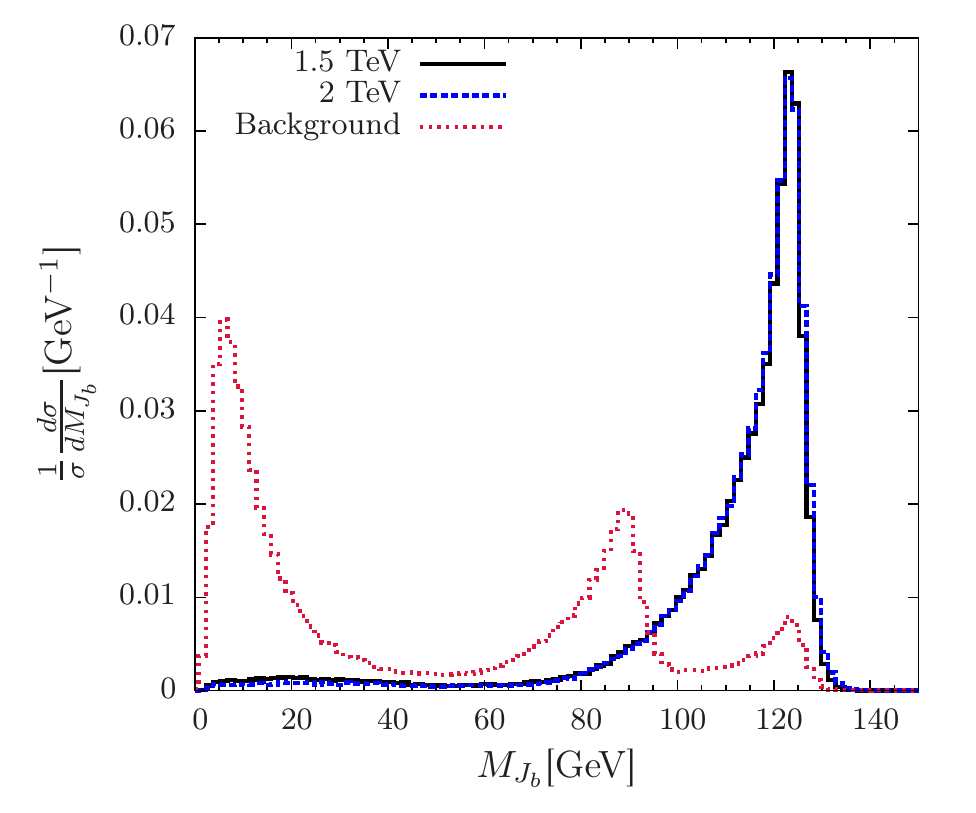}
\caption{Fat b-Jet mass $(M_{J_b})$ distribution from the signal and background events for $M_N=700$ GeV and $800$ GeV at the $\sqrt{s}=1$ TeV (left panel) and $M_N=1.5$ TeV and $2$ TeV at the $\sqrt{s}=3$ TeV (right panel) linear colliders.}
\label{ILC mfatb histogram}
\end{figure}

In Figs.\ref{ILC METfatb histogram}, \ref{ILC ptfatb histogram} and \ref{ILC mfatb histogram}, we plot the missing momentum $(p_T^{miss})$, transverse momentum of the fat-b jet $p_{T}^{J_{b}}$ and jet mass of the fat-b jet 
$(M_{J_{b}})$ distributions for $M_{N}=700$ GeV and $800$ GeV at the $\sqrt{s}=1$ TeV linear collider and $M_{N}=1.5$ TeV and $2$ TeV at the $\sqrt{s}=3$ TeV linear collider. 
In view of these distributions, we have used the following advanced selection cuts to reduce the SM background:

\subsubsection{Advanced cuts for $M_{N}=400$ GeV- $900$ GeV at the $\sqrt{s}=1$ TeV linear collider after the detector simulation}
\begin{itemize}
\item Transverse momentum for  $J_b$, $p_{T}^{J_b}>250$ GeV.
\item Fat-b mass, $M_{J_b}>115$ GeV.
\item Missing energy, $p_{T}^{miss}>150$ GeV.
\end{itemize}

We consider two benchmark points such as $M_N=700$ GeV and $800$ GeV at the $1$ TeV linear collider to produce the boosted Higgs from RHNs. The cut flow has been shown in Tab.~\ref{ILCHiggs1TeV}. The b-jets are coming from the SM $h$ as the $M_{J_b}$ distribution peaks at the Higgs mass for the signal at the linear colliders. As a result $M_{J_b} > 115$ GeV sets a strong cut on the SM backgrounds.
\begin{table*}[!htbp]
 \begin{tabular}{|c|c|c|c|}
  \hline
  Cuts & \multicolumn{2}{|c|}{Signal}  & Background \\ \hline
       & $M_{N}=700$ GeV & $M_{N}=800$ GeV    &             \\    
  \hline
  Basic Cuts & 1,288,150 & 1,248,340 & 19,300 \\   
   $p_{T}^{miss}>150$ GeV & 1,239,440 & 1,223,480 & 8,373 \\
  $p_{T}^{J_{b}}>250$ GeV & 1,100,790 & 1,153,650 & 4,239 \\
  $M_{J_{b}}>115$ GeV & 609,330 & 661,258 & 855 \\ 
  \hline
 \end{tabular}
\caption{Cut flow for the signal and background events for the final state $J_b+ p_T^{miss}$ for $M_{N}=700$ GeV and $800$ GeV at the $\sqrt{s}=1$ TeV linear collider. The signal events are normalized by the square of the mixing. }
\label{ILCHiggs1TeV}
\end{table*}
\subsubsection{Advanced cuts for the $M_{N}=1$ TeV -$2.9$ TeV for the $\sqrt{s}=3$ TeV linear collider after the detector simulation}
\begin{itemize}
\item Transverse momentum for fat-b $(J_b)$, $p_{T}^{J_b}>350$ GeV.
\item Fat-b mass, $M_{J_b}>115$ GeV.
\item Missing energy, $p_{T}^{miss}>175$ GeV.
\end{itemize}

We consider two benchmark points such as $M_N=1.5$ TeV and $2$ TeV at the $3$ TeV linear collider for the boosted Higgs production from the RHN. The cut flow has been shown in Tab.~\ref{ILCHiggs3TeV}. The b-jets are coming from the SM $h$ as the $M_{J_b}$ distribution peaks at the Higgs mass for the signal at the linear colliders. As a result $M_{J_b} > 115$ GeV sets a strong cut on the SM backgrounds. We also consider a strong $p_T^{J_b} > 350$ GeV cut for the high mass RHNs at the $3$ TeV collider.
\begin{table*}[!htbp]
 \begin{tabular}{|c|c|c|c|}
  \hline
   Cuts & \multicolumn{2}{|c|}{Signal}  & Background \\ \hline
       & $M_{N}=1.5$ TeV & $M_{N}=2$ TeV    &             \\    
  \hline
  Basic Cuts & 5,077,160 & 4,043,130 & 74,245 \\
  $p_{T}^{miss}>175$ GeV & 5,005,240 & 4,011,420 & 39,231    \\
  $p_{T}^{J_{b}}>350$ GeV & 4,731,550 & 3,902,490 & 15,327 \\
  $M_{J_{b}}>115$ GeV & 2,961,620 & 2,479,960 & 3,740 \\  
  \hline
 \end{tabular}
\caption{Cut flow for the signal and background events for the final state $J_b+ p_T^{miss}$ for $M_{N}=1.5$ TeV and $2$ TeV at the $\sqrt{s}=3$ TeV linear collider. The signal events are normalized by the square of the mixing.}
\label{ILCHiggs3TeV}
\end{table*}
In this work, we adopt a minimalistic approach and consider a flat $70\%$ tagging efficiency for each of the daughter b jets coming from the Higgs decay.
\section{Current bounds}
\label{sec4}
\begin{figure}[]
\centering
\includegraphics[width=0.95\textwidth]{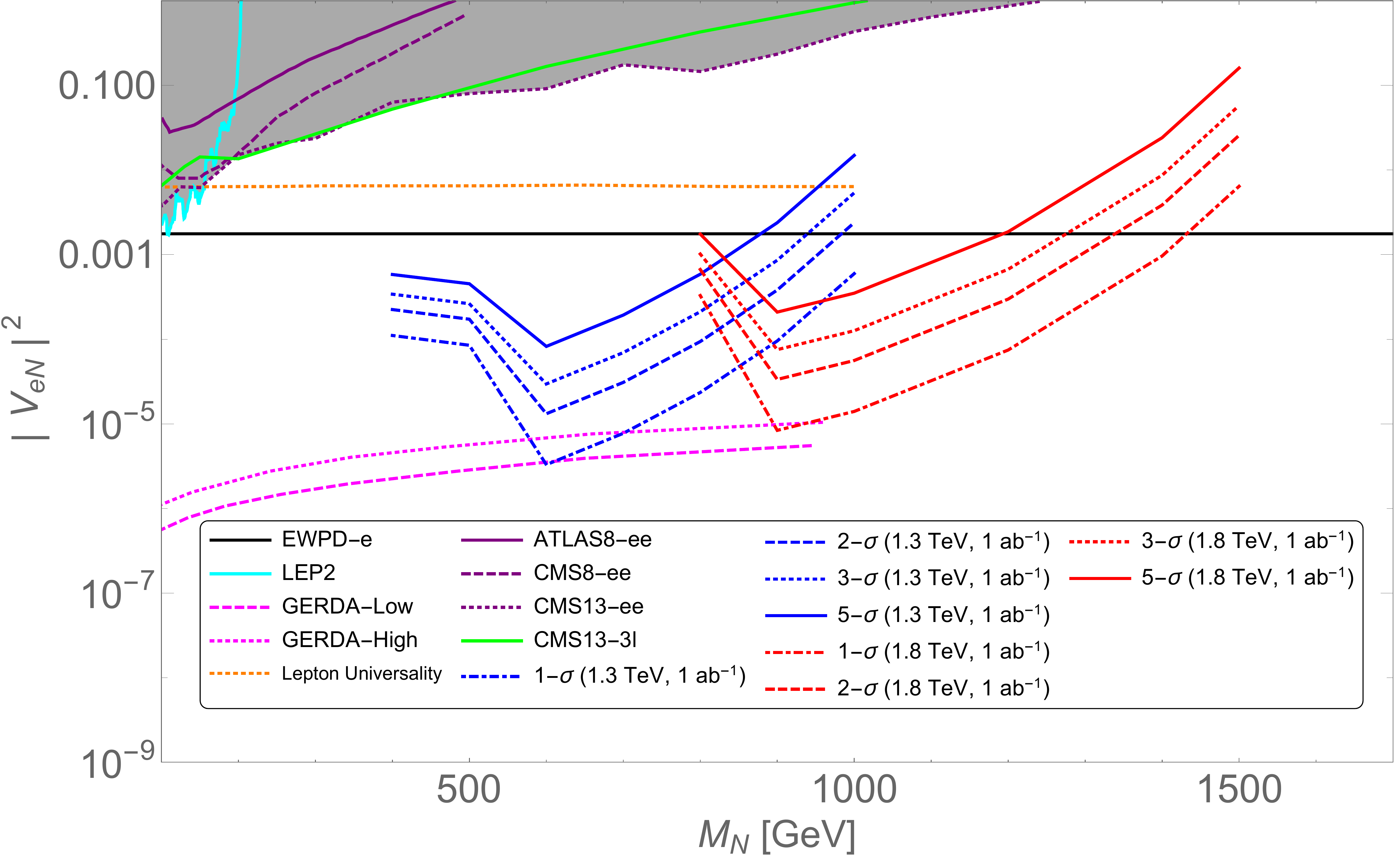}
\caption{The prospective upper limits on $|V_{eN}|^2$ at the $1.3$ TeV LHeC (blue band) and $1.8$ TeV HE-LHeC (red band) at the $1$ ab$^{-1}$ luminosity compared to EWPD \cite{delAguila:2008pw,Akhmedov:2013hec,deBlas:2013gla}, LEP2\cite{Achard:2001qv}, GERDA \cite{Agostini:2013mzu} $0\nu 2\beta$ study from \cite{Deppisch:2015qwa}, ATLAS (ATLAS$8$-$ee$) \cite{Aad:2015xaa}, CMS (CMS$8-ee$) \cite{Khachatryan:2016olu} at the $8$ TeV LHC, $13$ TeV CMS search for $e^\pm e^\pm+2j$ (CMS$13$-$ee$) \cite{Sirunyan:2018xiv} and $13$ TeV CMS search for $3\ell$ (CMS$13$-$ee$) \cite{Sirunyan:2018xiv} respectively.}
\label{Mix1}
\end{figure}

\begin{figure}[]
\centering
\includegraphics[width=0.95\textwidth]{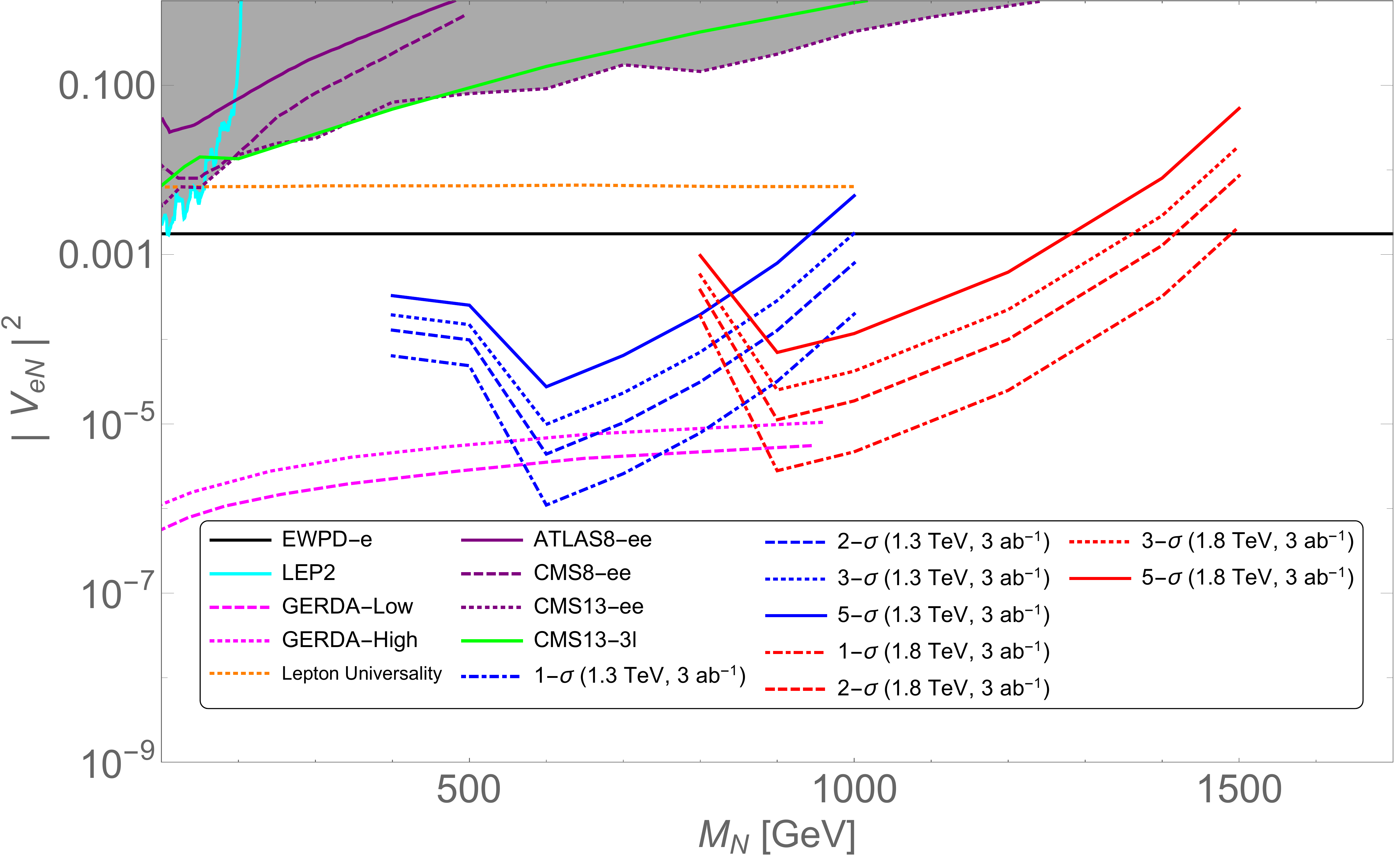}
\caption{Same as Fig.~\ref{Mix1} with $3$ ab$^{-1}$ luminosity at the $1.3$ TeV LHeC and $1.8$ TeV HE-LHeC.}
\label{Mix10}
\end{figure}

\begin{figure}[]
\centering
\includegraphics[width=0.95\textwidth]{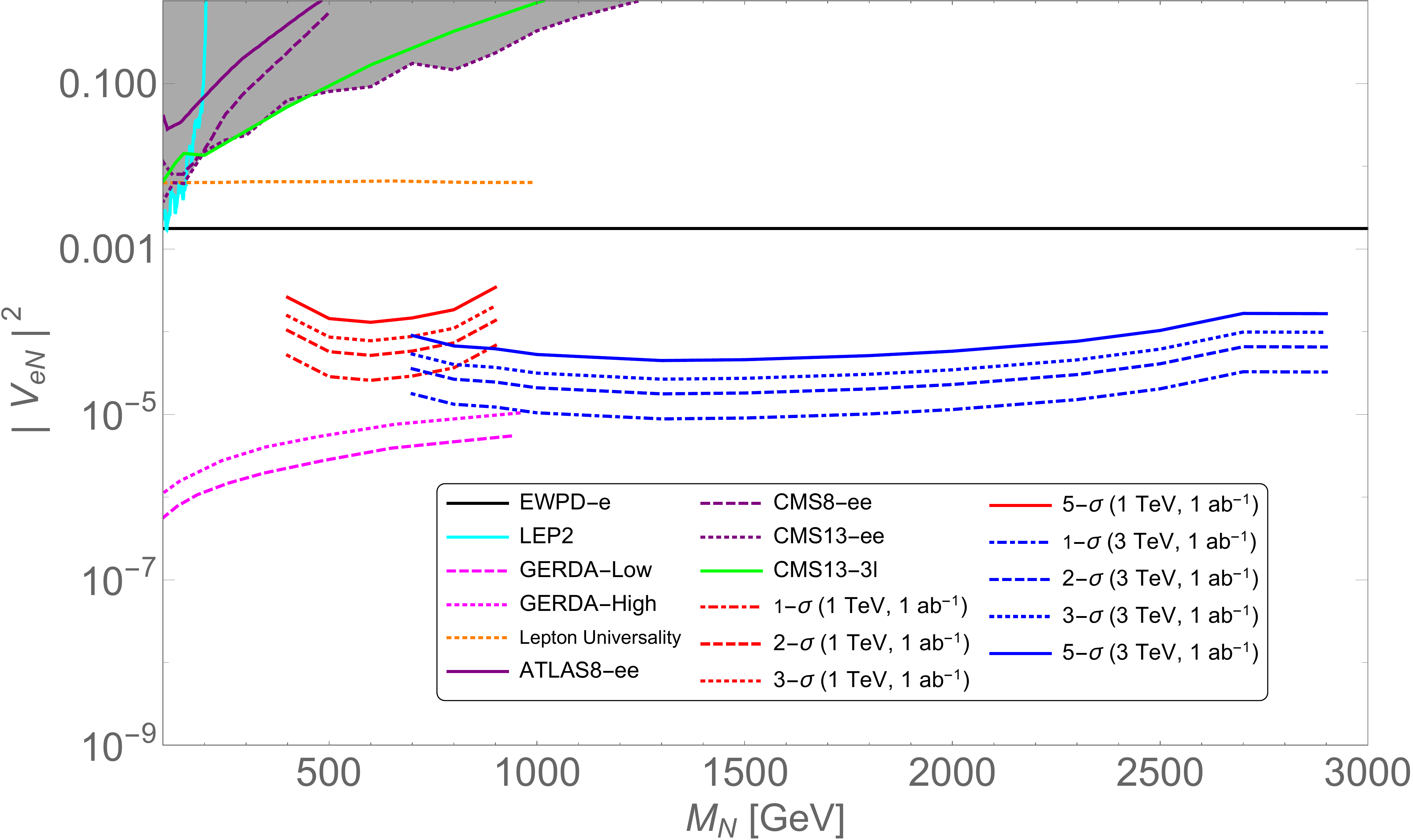}
\caption{The prospective upper limits on $|V_{eN}|^2$ at the $1$ TeV (red band) and $3$ TeV (blue band) linear colliders at the $1$ ab$^{-1}$ luminosity for $e+J+p_T^{miss}$ signal compared to EWPD \cite{delAguila:2008pw,Akhmedov:2013hec,deBlas:2013gla}, LEP2\cite{Achard:2001qv}, GERDA \cite{Agostini:2013mzu} $0\nu 2\beta$ study from \cite{Deppisch:2015qwa}, ATLAS (ATLAS$8$-$ee$) \cite{Aad:2015xaa}, CMS (CMS$8-ee$) \cite{Khachatryan:2016olu} at the $8$ TeV LHC, $13$ TeV CMS search for $e^\pm e^\pm+2j$ (CMS$13$-$ee$) \cite{Sirunyan:2018xiv} and $13$ TeV CMS search for $3\ell$ (CMS$13$-$ee$) \cite{Sirunyan:2018xiv} respectively.}
\label{Mix2}
\end{figure}

\begin{figure}[]
\centering
\includegraphics[width=0.95\textwidth]{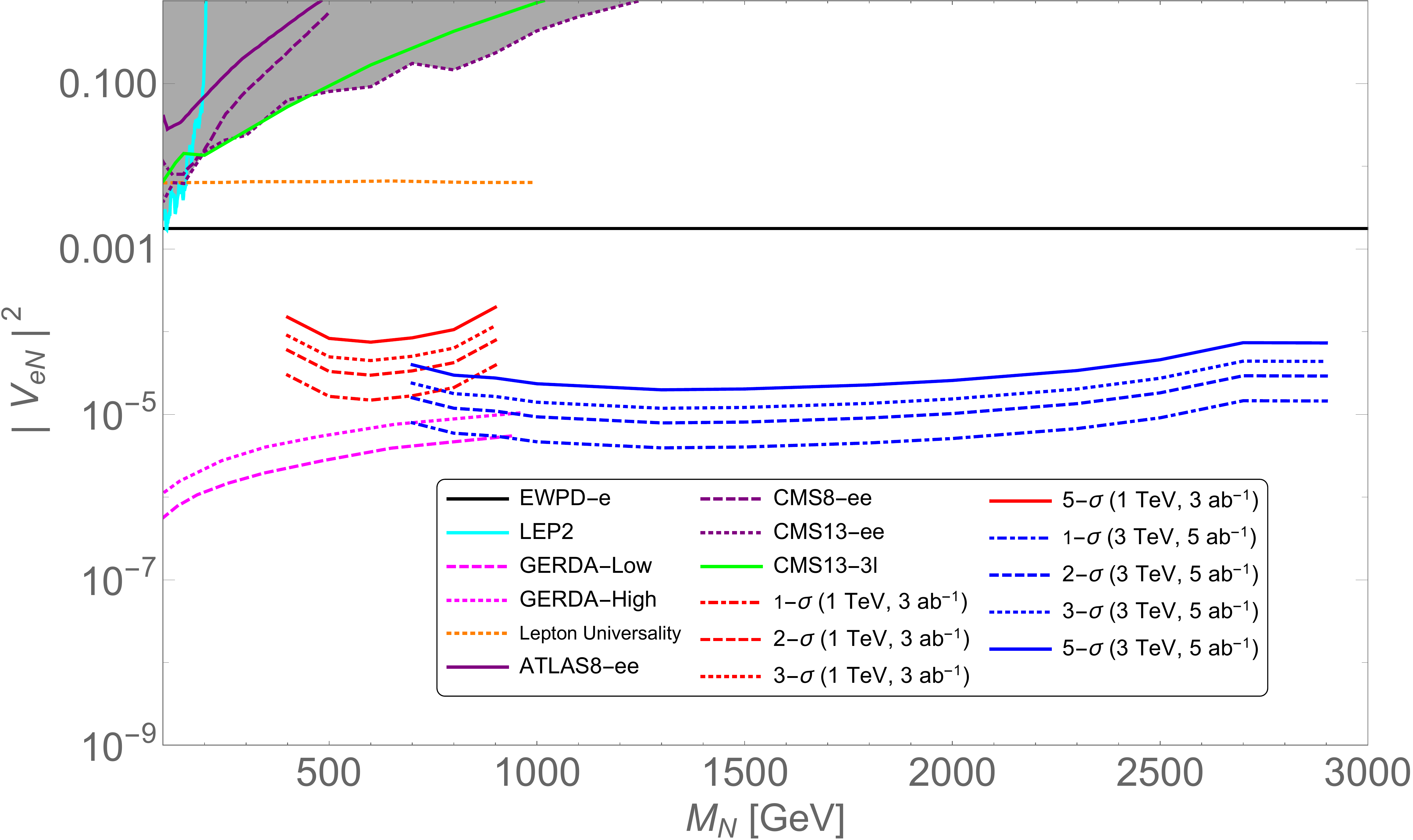}
\caption{Same as Fig.~\ref{Mix2} with $3(5)$ ab$^{-1}$ luminosity at the $1(3)$ TeV linear collider.}
\label{Mix20}
\end{figure}

\begin{figure}[]
\centering
\includegraphics[width=0.95\textwidth]{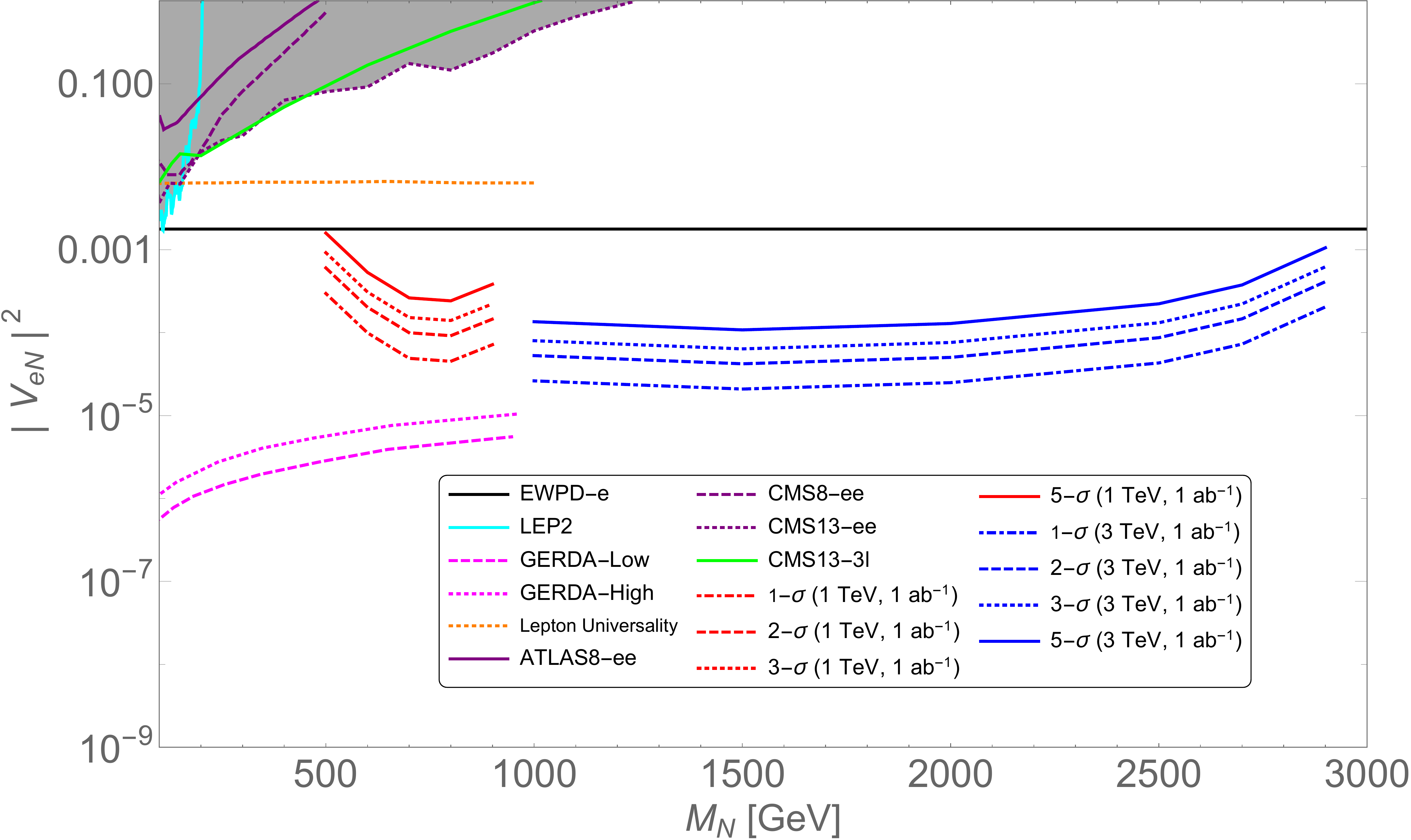}
\caption{The prospective upper limits on $|V_{eN}|^2$ at the $1$ TeV (red band) and $3$ TeV (blue band) linear colliders at the $1$ ab$^{-1}$ luminosity for $J_b+p_T^{miss}$ signal compared to EWPD \cite{delAguila:2008pw,Akhmedov:2013hec,deBlas:2013gla}, LEP2\cite{Achard:2001qv}, GERDA \cite{Agostini:2013mzu} $0\nu 2\beta$ study from \cite{Deppisch:2015qwa}, ATLAS (ATLAS$8$-$ee$) \cite{Aad:2015xaa}, CMS (CMS$8-ee$) \cite{Khachatryan:2016olu} at the $8$ TeV LHC, $13$ TeV CMS search for $e^\pm e^\pm+2j$ (CMS$13$-$ee$) \cite{Sirunyan:2018xiv} and $13$ TeV CMS search for $3\ell$ (CMS$13$-$ee$) \cite{Sirunyan:2018xiv} respectively.}
\label{Mix3}
\end{figure}

\begin{figure}[]
\centering
\includegraphics[width=0.95\textwidth]{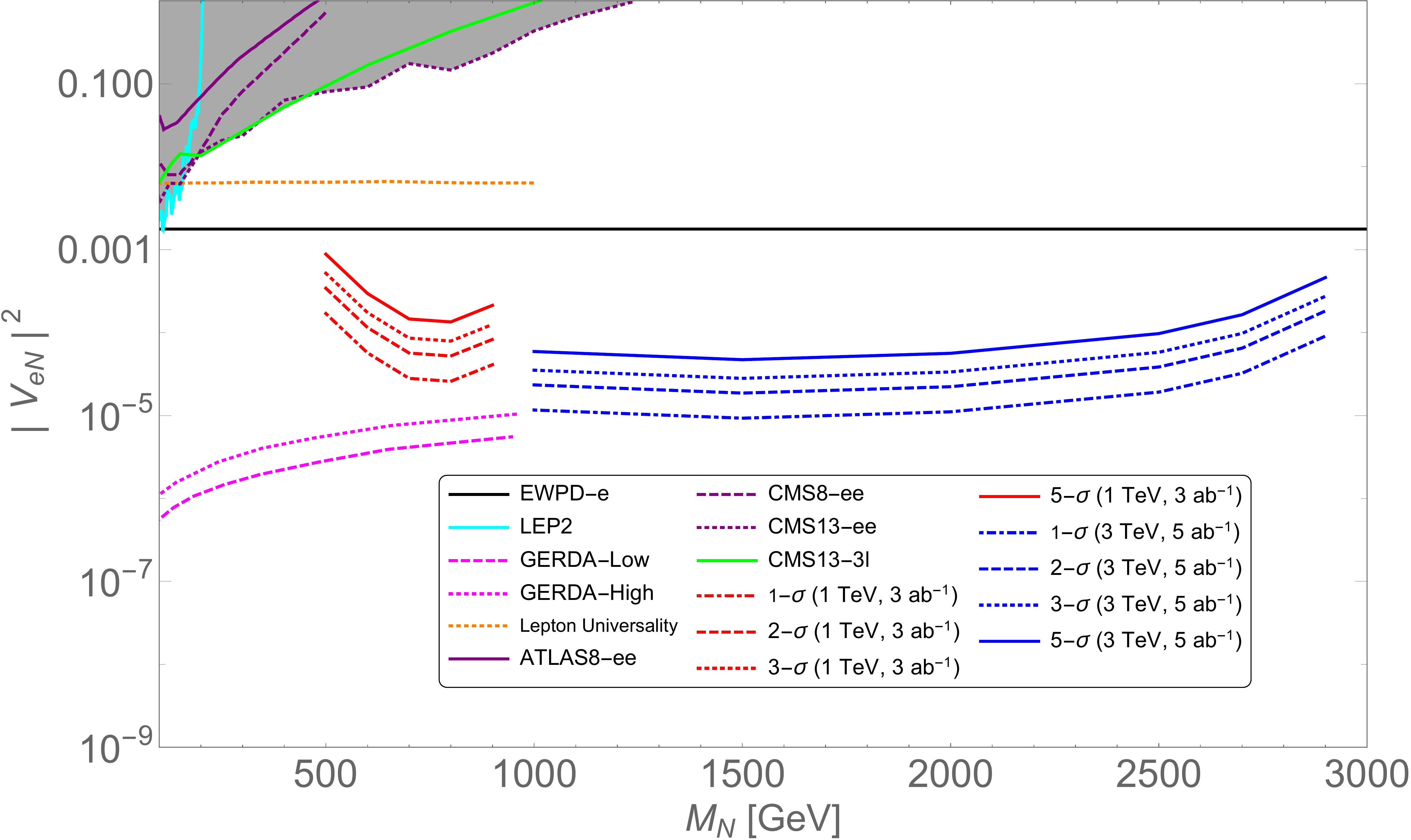}
\caption{Same as Fig.~\ref{Mix3} with $3(5)$ ab$^{-1}$ luminosity at the $1(3)$ TeV linear collider.}
\label{Mix30}
\end{figure}

The bounds on the light-heavy neutrino mixing for the electron flavor comes from a variety of searches. As we are interested on the RHN of mass $M_N \geq 100$ GeV, therefore we will compare our results with such bounds which are important for that mass range.  The Electroweak Precision Data (EWPD) bounds have been calculated in \cite{delAguila:2008pw,Akhmedov:2013hec,deBlas:2013gla} which obtains the bound on $|V_{eN}|^2$ as $1.681\times 10^{-3}$ at the $95\%$ C. L., the LEP2\cite{Achard:2001qv}, calculated at the $95\%$ C.L., bounds are rather weaker except $M_N=108$ GeV where it touches the EWPD line. The strongest bounds are coming from the GERDA \cite{Agostini:2013mzu} $0\nu 2\beta$ study where the limits as calculated in \cite{Deppisch:2015qwa} up to $M_N=959$ GeV. The lepton universality limits from \cite{deGouvea:2015euy} set bounds on $|V_{eN}|^2$ at $6.232\times 10^{-4}$ up to $M_N= 1$ TeV at the $95\%$ C. L. These bounds are plotted in Figs.~\ref{Mix1} -\ref{Mix30}.
 
Apart from the above mentioned indirect searches, the recent collider searches for the LHC also set bounds $|V_{eN}|^2$  at the $\sqrt{s}=8$ TeV at $95\%$ C. L. from same sign dilepton plus dijet search. The bounds on $|V_{eN}|^2$ from ATLAS (ATLAS$8$-$ee$) \cite{Aad:2015xaa} and CMS (CMS$8-ee$) \cite{Khachatryan:2016olu} are obtained at $23.3$ fb$^{-1}$ and $19.7$ fb$^{-1}$ luminosities respectively for the $e^\pm e^\pm+2j$ sample. The ATLAS limit is weaker than the CMS limits for $100$ GeV $\leq M_N \leq 500$ GeV. The LHC has also published the recent results at $\sqrt{s}=13$ TeV with $35.9$ fb$^{-1}$ luminosity which set stronger bounds on $|V_{eN}|^2$ than the previous direct searches for $100$ GeV $\leq M_N \leq 500$ GeV. The bounds on $|V_{eN}|^2$ from the $e^\pm e^\pm+2j$ signal in CMS (CMS$13$-$ee$) \cite{Sirunyan:2018xiv} and from trilepton search at CMS (CMS$13$-$3\ell$) \cite{Sirunyan:2018mtv} are also competitive, however, weaker than the EWPD for $100$ GeV $\leq M_N \leq 1.2$ TeV. These limits are also plotted in Figs.~\ref{Mix1} -\ref{Mix30}.

We have explored that at the LHeC with $\sqrt{s}=1.3$ TeV collider energy and $1$ ab$^{-1}$ luminosity, the bound on $|V_{eN}|^2$ for  $MN_N=600$ GeV with $1$-$\sigma$ C.L. is better than the $0\nu2\beta$ limit from GERDA-low where as  $M_N \geq 959$ GeV at $1$-$\sigma$ limit can be probed better than the GERDA-low and high limit \cite{Agostini:2013mzu,Deppisch:2015qwa}. The GERDA limits are stronger for the $M_N$ benchmarks we have studied. The results have been shown in Fig.~\ref{Mix1}. In the same figure we show the bounds obtained from the HE-LHeC with $\sqrt{s}=1.8$ TeV collider energy and $1$ ab $^{-1}$ luminosity. In this case the current GERDA bounds are stronger up to $M_N=959$ GeV \cite{Agostini:2013mzu, Deppisch:2015qwa}. At the HE-LHeC RHN up to $M_N=1.2$ TeV can be probed at $5$-$\sigma$ and these bounds could be stronger than the limits obtained from the EWPD-e limit \cite{delAguila:2008pw,Akhmedov:2013hec,deBlas:2013gla}. The improved scenario at the $3$ ab$^{-1}$ luminosity for the LHeC and HE-LHeC are shown in Figs.~\ref{Mix10}.

At the linear collider we have explored two sets of signals. one is the $e+J+p_T^{miss}$ and the other one is $J_b+p_T^{miss}$. Using $e+J+p_T^{miss}$ signal at the $1$ TeV linear collider we have probed RHNs between $400$ GeV $\leq M_N \leq 900$ GeV at $5$-$\sigma$ but the $0\nu2\beta$ limit from GERDA \cite{Deppisch:2015qwa} is stronger than this result for $M_N \leq 959$ GeV, however, the bounds on $|V_{eN}|^2$ for the RHNs heavier than $1$ TeV can be probed at $5$-$\sigma$ significance or more at the linear collider with the $3$ TeV center of mass energy. All the results are. In this case apart from the fat jet properties, the polar angle cut for the leptons worked nicely. The results are shown in Fig.~\ref{Mix2}. We have also studied the linear colliders at $1(3)$ TeV center of mass energy with $3(5)$ ab$^{-1}$ luminosity. We can find the improved results in Fig.~\ref{Mix20}. Using the $J_b+p_T^{miss}$ signal we did a complementarity check where $M_N \geq 1$ TeV can be probed better than GERDA \cite{Deppisch:2015qwa} at $5$-$\sigma$ significance or more at the $3$ TeV linear collider. The linear collider can probe $|V_{eN}|^2$ down to $\mathcal{O}(10^{-5})$ for $M_N=1.35$ TeV at $3$ TeV, however, compared to this the bounds obtained at the $1$ TeV linear collider are weaker. The corresponding bounds at the $\sqrt{s}=1$ TeV and $3$ TeV linear collider are plotted in Figs.~\ref{Mix2} and \ref{Mix3}. The red (blue) band represents the bounds on $|V_{eN}|^2$ at $1$ TeV ($3$ TeV) linear collider at different confidence levels. Comparing the bounds between the final states $e+J+p_T^{miss}$ and $J_b+p_T^{miss}$ we find that the former one puts slightly stronger limits on $|V_{eN}|^2$. The results are shown in Fig.~\ref{Mix3}. We have also studied the linear colliders at $1(3)$ TeV center of mass energy with $3(5)$ ab$^{-1}$ luminosity. We can find the improved results in Fig.~\ref{Mix30}. Finally we comment that our results at the linear collider are stronger than the limits obtained from the EWPD-e \cite{delAguila:2008pw,Akhmedov:2013hec,deBlas:2013gla} throughout the study.

\section{Conclusion}
\label{sec5}
We have studied the RHNs which can be responsible for the generation of the tiny light neutrino masses. We have calculated the production cross sections for the RHNs at the LHeC and linear collider at various center of mass energies and followed by that we  have tested the discovery prospects of this RHNs. We have chosen $\sqrt{s}=1.3$ TeV and $1.8$ TeV for the LHeC and $\sqrt{s}=1$ TeV and $3$ TeV for the linear collider. We have considered the sufficiently heavy mass range of the RHNs. These RHNs can decay dominantly into $\ell W$ mode. A massive RHN can sufficiently boost the $W$ such that its hadronic decay modes can form a fat-jet. Therefore we study $e+j_1+J$ and $e+J+p_T^{miss}$ at the LHeC and linear collider respectively. Similarly we consider another interesting mode $N\to h \nu, h \to b\overline{b}$ where a boosted SM Higgs can produce a fat b-jet and test the $J_b+p_T^{miss}$ final state at the linear collider. Simulating the events and passing through the selection cuts for the different colliders we calculate the bounds on $|V_{eN}|^2$ at different luminosities and compare with the existing bounds. Hence we conclude that $M_N \geq 959$ GeV can be successfully probed at the $1.8$ TeV at the at $5$-$\sigma$ C. L. with $1$ ab$^{-1}$ and $3$ ab$^{-1}$ luminosities respectively. Whereas $M_N \leq 2.9$ TeV can be probed at the $3$ TeV linear collider with more than $5$-$\sigma$ C.L using the $e+J+p_T^{miss}$ signal.  
A complementary signal of $J_b+p_T^{miss}$ can be useful, too but this is weaker than the bounds obtained by the $e+J+p_T^{miss}$ final state.

\textbf{Note added: }
While in final drafting phase, we noticed Ref.\cite{mani} appeared in arXiv which also studied fat jet signatures from RHNs at the linear colliders. We have studied LHeC and linear collider at different center of mass energies using detailed cut based analyses. We have compared our results with all the existing bounds using the decay modes of the RHNs to $W$ and SM $h$ bosons. The $0\nu2\beta$ bound became very strong up to $M_N=959$ GeV. At the linear collider the polar angle variable for the lepton became very useful for us. In our analysis we have showed that high mass RHNs can be observed at $5$-$\sigma$ significance or more in these colliders. 
 \section*{Acknowledgement}
The work of SN and SJ was in part supported by US Department of Energy Grant Number DE-SC 0016013. AD would like to thank Oliver Fischer for useful discussions on the LHeC and sharing the DELPHES card for that detector.
AD would also like to thank Daniel Jeans (KEK), Kentarou Mawatari (Osaka University) and Junping Tian (KEK) for useful information regarding the linear collider card in DELPHES. 

\end{document}